\documentclass[twocolumn,pra]{revtex4-1}	
\usepackage{float}		
\usepackage{times}		
\usepackage{graphicx}	
\usepackage{url}
\usepackage{amsmath}
\usepackage{amsfonts}
\usepackage{amssymb}
\usepackage[mathscr]{euscript}
\usepackage{cancel}
\usepackage{color} 
\usepackage{hyperref}

\definecolor{grey}{rgb}{.6,.6,.6}

\def    \bse{\begin{subequations}}
\def    \ese{\end{subequations}}

\newcommand{\varg}{
\mathrm{\emph{g}}
}

\def\avg[#1]{
\langle #1 \rangle			
}

\begin{document}

\title{Large gain quantum-limited qubit measurement using a two-mode nonlinear cavity}
\author{S. Khan$^{\mathrm{1}}$, R. Vijay$^{\mathrm{2}}$, I. Siddiqi$^{\mathrm{3}}$, and A. A. Clerk$^{\mathrm{1}}$ \\ $^{\mathrm{1}}$\emph{Department of Physics, McGill
University, Montreal, Quebec, Canada H3A 2T8} \\ $^{\mathrm{2}}$\emph{Tata Institute of Fundamental Research,
Mumbai 400005, India} \\ $^{\mathrm{3}}$\emph{Department of Physics, University of California, Berkeley, CA 94720} }
\date{\today}

\begin{abstract}
We provide a thorough theoretical analysis of qubit state measurement in a setup where a driven, parametrically-coupled cavity system is directly coupled to the qubit, with one of the cavities having a weak Kerr nonlinearity.  Such a system could be readily realized using circuit QED architectures.  We demonstrate that this setup is capable in the standard linear-response regime of both producing a highly amplified output signal while at the same time achieving near quantum-limited performance: the measurement backaction on the qubit is near the minimal amount required by the uncertainty principle. This setup thus represents a promising route for performing efficient large-gain qubit measurement that is completely on-chip, and that does not rely on the use of circulators or complex non-reciprocal amplifiers.
\end{abstract}

\maketitle			

\section{Introduction}

Among the many attractive aspects of circuit QED approaches to quantum information processing is the ability to use driven microwave cavities for high-fidelity qubit measurements \cite{Blais04,Schuster05,Majer07}. The standard approach is  to couple the qubit dispersively to the cavity, such that the resonant frequency of the cavity depends on the state of the qubit.  By driving the cavity, the state of the qubit becomes encoded in the phase of the reflected drive tone. For an ideal realization, this initial measurement is quantum limited:  the backaction disturbance to the qubit is as  small as is permitted by quantum mechanics \cite{Blais04,ClerkRMP}.  In practice though, the qubit signal in the cavity output is too small to be sent directly to a room-temperature amplifier.  One thus often first uses a low temperature amplification stage, based on a Josephson parametric amplifier (JPA) \cite{Yurke87, Yurke89, Beltran07, Lehnert08, Nakamura08, Bergeal2010, Abdo11, Roch12, Siddiqi13}.  This amplifier is yet another driven microwave cavity, one that is made nonlinear by the addition of Josephson junctions.  While such setups  can yield extremely fast measurement rates, they unfortunately require the use of lossy magnetic circulators.  This both adds to experimental complexity and bulkiness, and makes one susceptible to insertion losses.

A promising alternate approach is to combine the measurement and amplification cavities in the standard setup into a single system.  One thus has an intrinsically nonlinear cavity which is directly coupled to a qubit, eliminating the need for circulators.  Such approaches have been studied in regimes where the nonlinear cavity is driven past a bifurcation point in its dynamics, yielding the Josephson bifurcation amplifier \cite{Siddiqi04, Siddiqi05, Lupascu06, Vijay09}, and a strong latching-style measurement of the qubit state.  To obtain a continuous measurement, one could drive the nonlinear cavity close to (but not past) a point of bifurcation, similar to the driving conditions used to realize a paramp.  Setups of this kind have been studied both theoretically \cite{Esteve10,Laflamme2011,Boissonneault2012,Laflamme2012,Boissonneault2014} and in experiment \cite{Esteve10,Vijay10b, Johnson2011}.  While the nonlinearity in these systems allows for large gain, this comes at the expense of
an unavoidable large excess backaction on the qubit~\cite{Laflamme2011}.  While this problem can be mitigated in more complex regimes having a stronger qubit-cavity coupling~\cite{Laflamme2012}, for a single nonlinear cavity weakly coupled to a qubit, this excess backaction prevents quantum-limited performance.

\begin{figure}[t]
\includegraphics[scale = 0.275]{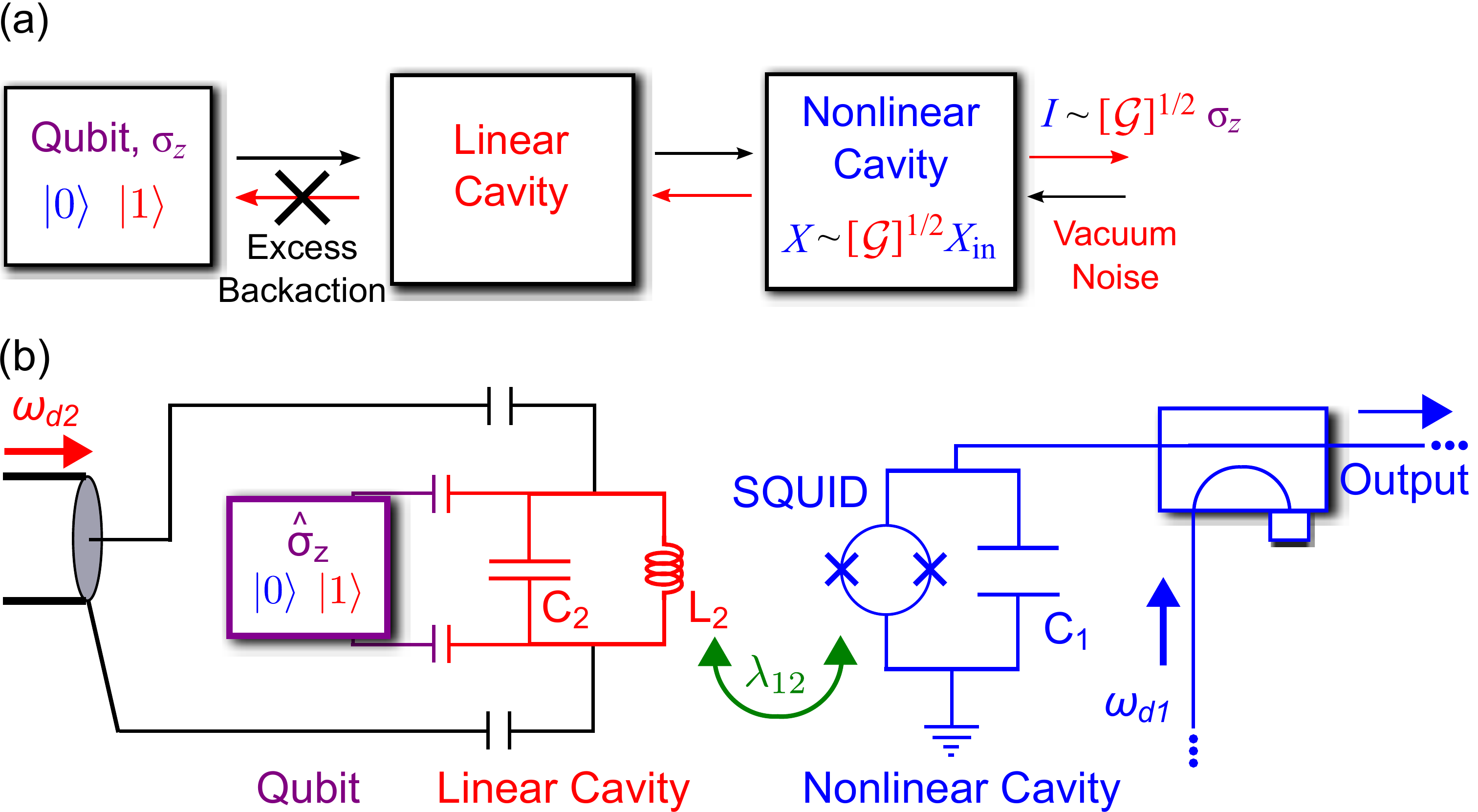}
\caption{(a)  Schematic diagram of the PARI setup. The linear cavity is used to provide tunable coupling between the qubit $\hat{\sigma}_z$ and the nonlinear cavity; it can be adjusted to suppress excess backaction, while still allowing the output homodyne current $\hat{I}$ to see the amplified quadrature $\hat{X}_e$ (cf. Eq.~(\ref{Xe})) and therefore the large gain $\mathcal{G}$ (cf. Eq.~(\ref{parGain})). (b) Possible experimental realisation of the PARI setup in (a) in a superconducting circuit architecture. Both cavities are driven, at frequencies $\omega_{dj}$ for $j=(1,2)$, and $\lambda_{12}$ is the intercavity coupling strength. All components shown above can be cofabricated on-chip as an integrated qubit-amplifier system.}
\label{boxSchematic}
\end{figure}

In this paper, we consider theoretically a new approach to making continuous qubit measurements.  The goal is again to combine the measurement cavity and paramp cavities in a traditional setup into a single, circulator-free device.  Unlike the work described above, we now consider directly coupling a qubit to a simple {\it multi-mode} structure with intrinsic nonlinearity.  Our basic setup is shown in Fig.~\ref{boxSchematic}. Similar to a standard dispersive-measurement setup, we have a driven linear cavity dispersively coupled to a qubit, as well as a second driven nonlinear cavity.  However, unlike the standard setup, we do \emph{not} take the output field of the linear cavity and use it to drive (via a circulator) the nonlinear cavity.  We instead {\it directly couple} these systems, such that the intracavity field of the linear cavity directly modulates the inductance of the nonlinear cavity. We refer to this setup as a ``parametric amplifier with resonant input'' (PARI). Note that the amplification properties of a very different kind of two-mode nonlinear cavity were recently studied by Eichler \emph{et al.} in Ref.~\onlinecite{eichler2014}; unlike our work, they did not consider a direct coupling to a qubit.

It would appear that the PARI will suffer the same fate encountered when directly coupling a qubit to a single nonlinear cavity:  there is nothing to protect the qubit from the large fluctuations associated with the nonlinear cavity, hence there will be backaction on the qubit far in excess of the quantum limit.  Protecting the qubit from such fluctuations is precisely the role of the circulator in a standard cavity-circulator-paramp setup; the directional nature of the circulator prevents amplified noisy signals emanating from the amplifier input port from reaching the qubit. Similar protection could be achieved without a circulator, if one had a truly directional amplifier, one that did not produce amplified noise at its input port~\cite{Devoret2014}.  

In the PARI design, there is no explicitly directional element preventing the qubit from seeing the nonlinear cavity and its large, amplified noise. The only element mitigating this coupling is the linear cavity.  As we will see, {\it this is already enough to effectively protect the qubit from excess backaction}. There is an emergent directionality here as a result of having a driven system, and having the freedom to control the relative phase between the drive tones applied to each of the two cavities. We stress that the linear cavity does not act as a simple filter, suppressing noise at high/low frequencies.  Rather, the driven linear cavity acts to control the phase-sensitive coupling of the qubit to the nonlinear cavity, in a manner that would be impossible if the qubit was coupled directly to the nonlinear cavity (as in, e.g., Ref.~\cite{Laflamme2011}).

Of course, simply protecting the qubit from backaction is not enough.  We also want the system to truly act as an amplifier, producing a sufficiently large qubit signal as well as a sufficiently short measurement time $\tau_{\rm meas} = 1 / \Gamma_{\rm meas}$ (i.e.~how long one must run the measurement before being able to distinguish the qubit ground state from the excited state).  Our analysis shows that for experimentally feasible parameters, the PARI system has enough flexibility to allow both these goals to be achieved while at the same time having near-ideal backaction properties.   We stress that as it contains no circulators, all components of our detector can be cofabricated on-chip.  Hence the PARI setup realizes an integrated qubit-amplifier system with a large output signal, fast measurement rate and near-quantum-limited backaction.

The remainder of this paper is organized as follows.  In Sec.~\ref{hamiltonians}, we introduce the model for the PARI system, and outline our basic analysis of the corresponding Heisenberg-Langevin equations; of particular note is Sec.~\ref{subsec:Mapping}, where we develop an extremely useful mapping of this two-mode system to a single effective nonlinear cavity.  In Sec~\ref{sec:results}, we turn to analyzing and optimizing the properties of the PARI detector, viewed as a qubit state detector.  We use linear-response theory to calculate the detector gain and noise properties. Our analysis reveals that optimal performance is possible using realistically weak cavity-cavity couplings, if one designs the system so that the damping rate of the nonlinear cavity is greater than that of the linear cavity, and weak nonlinearities are used.  Using realistic experimental parameters, we predict that near quantum-limited performance should be possible with photon number gains $\sim 20$~dB (cf. Fig.~\ref{GainVsDelta1}), and with fast measurement times $\sim 100~{\rm ns}$ (cf. Fig.~\ref{MeasVsDelta1}). The backaction is within a factor of $2$ of its quantum limited value:
 \begin{equation}
 	1/\eta_{\rm meas} = \Gamma_{\varphi} /  \Gamma_{\rm meas}  \simeq 1.75
\end{equation}
where $\Gamma_{\varphi}$ is the rate of backaction-induced qubit dephasing (directly proportional to the backaction noise driving the qubit, cf.~Eq.~(\ref{gammaPhi})), and $\Gamma_{\rm meas}$ is the measurement rate (inversely proportional to the detector imprecision noise, cf.~Eq.~(\ref{eq:GammaMeas})).  As discussed explicitly in section~\ref{compareConvention}, this performance (achieved completely on-chip) is on par with or better than modern circulator-based setups~\cite{Vijay2012, Murch2013a, Murch2013b, deLange2013}. In contrast, if one directly (and weakly) coupled a qubit to a single nonlinear cavity, a gain of $20$~dB would imply missing the quantum limit by a large factor of $\sim 50$.

\section{Basic two-cavity system and analysis}
\label{hamiltonians}
\subsection{Hamiltonian and coupling}

The two-cavity detector system we consider consists of a single nonlinear cavity  with a Kerr type nonlinearity, coherently coupled to an ordinary linear cavity. Throughout this paper, we label the nonlinear cavity with the index $1$, and the linear cavity with index $2$. The two cavity Hamiltonians are given in terms of the cavity annihilation operators $\hat{a}_j$ ($j=1,2$) by
\begin{subequations}\label{grp}
\begin{align}
\hat{H}_1 = \hat{H}^{\mathrm{sys}}_1 + \hat{H}^{\mathrm{envt}}_1 &=  \omega_{c1}\hat{a}^{\dagger}_1\hat{a}_1 - \Lambda
\hat{a}^{\dagger}_1\hat{a}^{\dagger}_1 \hat{a}_1\hat{a}_1 + \hat{H}^{\mathrm{envt}}_1
\label{H1} \\
\hat{H}_2 = \hat{H}^{\mathrm{sys}}_2 + \hat{H}^{\mathrm{envt}}_2 &= \omega_{c2} \hat{a}^{\dagger}_2\hat{a}_2 +
 \hat{H}^{\mathrm{envt}}_2
\label{H2}
\end{align}
\end{subequations}
($\hbar = 1$) where $\omega_{cj}$ are the cavity resonance frequencies.  For the nonlinear cavity Hamiltonian [cf. Eq.~(\ref{H1})], the Kerr type nonlinearity has strength $\Lambda \ll \kappa_1$, where $\kappa_1$ is the nonlinear cavity damping rate; this system could be realised as a SQUID shunted with a capacitor in superconducting circuit architecture~\cite{Hatridge11}. We assume a regime where the phase fluctuations of the SQUID are small enough for only the leading order Kerr nonlinearity to be important; the conditions under which higher order terms in an expansion of the SQUID cosine potential can be dropped are discussed in Sec.~\ref{optimalPerformance}. Both cavities are also driven externally, via coupling to external microwave transmission lines. The $\hat{H}^{\mathrm{envt}}_j$ terms describe this driving, and the damping (at rate $\kappa_j$) of both cavities in a standard input-output theory approach~\cite{Gardiner00,ClerkRMP}; we neglect internal cavity losses.

The qubit to be measured is connected to the two-cavity system via a dispersive coupling to the \emph{linear} cavity~\cite{Blais04}; this coupling is described by the Hamiltonian,
\begin{equation}
	\hat{H}_{\sigma_z} =  A\hat{\sigma}_z\hat{a}^{\dagger}_2\hat{a}_2 \equiv \hat{\sigma}_z \hat{F},
	\label{Hsigmaz}
\end{equation}
where $\hat{\sigma}_z$ commutes with the qubit Hamiltonian. The cavity-$2$ resonance frequency is modulated by the qubit state by an amount $A$, the dispersive coupling strength. We have introduced $\hat{F}$, the effective backaction noise operator.  Throughout this paper we will consider values of $A$ small enough that a linear response analysis of the detector is valid; when considering real experimental parameters, we will explicitly check the validity of this assumption. As is standard with dispersive qubit measurements, we measure the qubit in the energy eigenstate ($\hat{\sigma}_z$) basis.  Measurements along other axes can be achieved by first performing a qubit rotation, as has been done in countless circuit QED experiments.

Finally, we describe the crucial element of the setup: how the two cavities are coupled to each other to allow the nonlinear cavity to learn about the qubit state. For the two-cavity system within superconducting circuit architecture, the qubit state-dependent current in cavity-$2$ generates a flux that threads the SQUID loop in cavity-$1$, therefore modulating its Josephson energy. This inductive coupling between the cavities takes the nonlinear form [see Appendix~\ref{couplingDerivation}],
\begin{equation}
\hat{H}_{12} = \frac{1}{2} \left[ \lambda_{12} (\hat{a}_2 + \hat{a}_2^{\dagger}) \right] (\hat{a}_1 + \hat{a}_1^{\dagger})^2
\label{H12}
\end{equation}
where $\lambda_{12}$ is the strength of the intercavity coupling. Note that the cavity-$2$ ``sum'' quadrature effectively modulates the `spring constant' of cavity-$1$.  

We now make an important choice for the two cavity drive frequencies, $\omega_{dj}~(j=1,2)$, taking the linear cavity to be driven at twice the driving frequency of the
nonlinear cavity:
\begin{equation}
\omega_{d2} = 2\omega_{d1}
\label{driveCond}
\end{equation}
This choice of frequencies is unusual, in that the nonlinear cavity is effectively being coupled to a higher frequency signal, the field of the linear cavity. In particular, as we will show in Eq.~(\ref{H12d}), this choice (along with a rotating-wave approximation) leads to an emergence of effective parametric driving and photon hopping terms in the coupling Hamiltonian, as opposed to the usual dispersive ($\hat{a}^{\dagger}\hat{a}$) terms (which are non-resonant in this case).

The full system Hamiltonian is then given by $\hat{H} = \hat{H}_1 + \hat{H}_2 + \hat{H}_{12} + \hat{H}_{\sigma_z}$. We emphasise that the results of our analysis hold for a generic Hamiltonian of the form of $\hat{H}$, involving a dispersive coupling between two independent modes. As mentioned earlier, our emphasis will be on weak qubit-detector couplings, allowing the use of quantum linear-response theory \cite{ClerkRMP}.  In what follows, we will thus start by developing a description of the PARI in the absence of any coupling to a qubit (i.e.~$A \rightarrow 0$), and will then use this description combined with linear-response theory to determine its properties as a detector.  

\subsection{Heisenberg-Langevin equations}

Treating the cavity driving and dissipation as in standard input-output theory \cite{Gardiner00,ClerkRMP}, the Heisenberg-Langevin equations for the two cavity modes are 
($j=1,2$)
\begin{equation}
\frac{d}{dt}\hat{a}_j = -i [\hat{a}_j,\hat{H}^{\mathrm{sys}}_j] - i
[\hat{a}_j,\hat{H}_{12}] -\frac{\kappa_j}{2}\hat{a}_j - \sqrt{\kappa_j}~\hat{a}^{\mathrm{in}}_j(t), 
\label{heisenbergA}
\end{equation}
where the driving fields describe both coherent drive tones (amplitudes $\bar{a}^{\mathrm{in}}_j$)  as well as noise:
\begin{align}
\hat{a}^{\mathrm{in}}_1 (t) &= \bar{a}^{\mathrm{in}}_1 e^{-i\omega_{d1}t} + \hat{\xi}_1(t), \nonumber \\
\hat{a}^{\mathrm{in}}_2 (t) &= \bar{a}^{\mathrm{in}}_2e^{i\delta} e^{-i\omega_{d2}t} + \hat{\xi}_2(t).
\label{drives}
\end{align}
Here,  $\omega_{dj} = \omega_{cj} + \Delta_j$, with $\Delta_j$ denoting the detuning from cavity resonance. As we will see, the relative phase between the cavity drive tones will play an important role in allowing one to tune the system properties. We denote this phase difference $\delta$, and without loss of generality take the $\bar{a}^{\mathrm{in}}_j$ in Eqs.~(\ref{drives}) to be real and positive. The operators $\hat{\xi}_j(t)$ describe the classical and quantum vacuum noises that enter each cavity via the drive port; the noise is taken to be delta-correlated vacuum noise in the standard input-output theory approach~\cite{Gardiner00,ClerkRMP}.

\subsection{Classical Cavity Amplitudes}

As a result of the strong coherent drives on both cavities, the average cavity field amplitudes will be non-zero, $\avg[\hat{a}_j] \neq 0$. Our first task will be to find these amplitudes by solving the classical equations of motion for our system.  Consistent with our focus on weak qubit-detector couplings and use of linear response, we will solve for these amplitudes in the absence of any coupling to the qubit, and then use them to transform to a displaced interaction picture (with respect to the cavity drive frequencies):
\begin{align}
\hat{a}_j(t) &= e^{-i\omega_{dj}t} \left\{\alpha_j + e^{i\mu_j}\left[ e^{i\beta}\hat{d}_j(t) \right] \right\} \,\,\,\,\,\,\,\,\, (j=1,2), \nonumber \\
\alpha_j &\equiv \avg[\hat{a}_j] \equiv e^{i\mu_j}\sqrt{\bar{n}_j} .
\label{modes}
\end{align}
Here $\bar{n}_j$ are the respective cavity photon numbers due to the classical drives, $\mu_j$ are the average cavity mode phases, and $\beta$ is an overall phase that will be chosen later for convenience. 

The values of the classical cavity mode amplitudes $\alpha_j$ are determined by solving the classical equations of motion for  both cavities (in the absence of any qubit coupling). The two coupled nonlinear equations obtained can be reduced to a single equation for the nonlinear cavity mode amplitude [see Appendix~\ref{appendixClassical}],
\begin{equation}
	\alpha_1 [ -i(\Delta_1 + 2\widetilde{\Lambda} |\alpha_1|^2) + \kappa_1/2] - i\eta\alpha_1^*= -\sqrt{\kappa_1}~\bar{a}^{\mathrm{in}}_1.  
	\label{classicalA1}
\end{equation}
We see that the cavity-cavity coupling modifies the classical amplitude equation for $\alpha_1$ from that of a standard Kerr cavity in two ways.  The first is a modification of the bare Kerr constant $\Lambda$ in Eq.~(\ref{H1}) to the complex valued $\widetilde{\Lambda}$,
\begin{equation}
	\widetilde{\Lambda} = \Lambda + \lambda_{12}^2 \left(\frac{i \chi_{2}}{4} \right),
\label{complexKerr}
\end{equation}
where $\chi_{2} \equiv \chi_{2}[\omega = 0]$ is the (complex) zero-frequency bare linear cavity susceptibility; for non-zero frequency,
\begin{equation}
 	\chi_{2}[\omega] = \left[ - i (\omega + \Delta_2) + \kappa_2/2  \right]^{-1} .
	\label{chi2}
\end{equation}
Also new, and more complicated, is the last term on the left hand side in Eq.~(\ref{classicalA1}), which involves $\alpha_1^*$; this
describes an effective parametric driving of cavity-$1$ due to the coupling to cavity-$2$. The effective parametric drive strength $\eta$ is given by
\begin{equation}
\eta =\lambda_{12}\left(\sqrt{\kappa_2}~\bar{a}^{\mathrm{in}}_2~\chi_2~e^{i\delta}\right).
\label{eta}
\end{equation}
It can be shown [see Appendix~\ref{appendixClassical}] that Eq.~(\ref{classicalA1}) can be written as a \emph{quintic} polynomial in the nonlinear cavity photon
number  $|\alpha_1|^2 = \bar{n}_1$~\cite{Wustmann13}:
\begin{align}
\bar{n}_1 = \left[ \frac{\kappa_1(\bar{a}_1^{\mathrm{in}})^2}{D^2} \right] &\left\{ \kappa_1/2 + i\eta + i(\Delta_1 +
2\widetilde{\Lambda}^*\bar{n}_1) \right\} \times \nonumber \\
&\left\{ \kappa_1/2 -i\eta^* -i(\Delta_1 + 2\widetilde{\Lambda}\bar{n}_1)  \right\}
\label{quinticClassical}
\end{align}
where
\begin{equation}
D = \left| \kappa_1/2 - i \left(\Delta_1 + 2\widetilde{\Lambda}\bar{n}_1 \right) \right|^2 - |\eta|^2
\end{equation}
 We see that in comparison to a  single driven Kerr cavity (where the average photon number is determined from a cubic polynomial), the two-mode structure of the PARI system yields more complex nonlinear physics at even the classical level.  In what follows, we will attempt to work in regimes where the cavity-cavity coupling $\lambda_{12}$ is weak enough that classical bifurcation physics described by Eq.~(\ref{quinticClassical}) is not too different from that of a single driven Kerr cavity. At a point of bifurcation, the driven nonlinear cavity becomes multistable when the cavity photon number $\bar{n}_1$ as a function of drive detuning acquires more than one stable solution. Just before this turning-over point is reached, the slope of $\bar{n}_1$ versus $\Delta_1$ is nearly vertical, and thus the cavity becomes very sensitive to even small changes in detuning; it is near (but not past) this point of bifurcation that we wish to operate.

\subsection{Linearization procedure}

Once the $\alpha_j$ are found, the displacement and interaction picture transformation in Eq.~(\ref{modes}) allows obtaining the transformed Heisenberg-Langevin equations of motion from Eqs.~(\ref{heisenbergA}):
\begin{equation}
\frac{d}{dt}\hat{d}_j = -i[\hat{d}_j,\hat{H}^d_j] - i[\hat{d}_j,\hat{H}^d_{12}] -
\frac{\kappa_j}{2}\hat{d}_j - \sqrt{\kappa_j}~\hat{\xi}_j(t), 
\label{heisenbergD}
\end{equation}
where $\hat{H}^d_j$ are the respective cavity Hamiltonians in the new frame.  We focus on the standard regime of strong driving and weak intrinsic nonlinearity ($\Lambda \ll \kappa_1$), which allows us to drop terms that are cubic and quartic in $\hat{d}_1, \hat{d}_1^\dagger$; formally, we are neglecting terms that are suppressed by factors of $1 / \bar{n}_1$. The nonlinear cavity Hamiltonian in the displaced interaction picture thus takes the form of a detuned degenerate parametric amplifier:
\begin{equation}
\hat{H}^d_1 = -\widetilde{\Delta}_1 \hat{d}_1^{\dagger}\hat{d}_1 -  \frac{\varg_1}{2}
(e^{-i2\beta}\hat{d}_1^{\dagger}\hat{d}_1^{\dagger} + h.c.) ,
\label{H1d}
\end{equation}
where the effective detuning $\widetilde{\Delta}_1$ and parametric interaction strengths $\varg_1$ are given by
\begin{subequations}\label{grp}
\begin{align}
\widetilde{\Delta}_1 &= \Delta_1 + 4\Lambda\bar{n}_1 
\label{D1E}  \\
\varg_1 &= 2\Lambda\bar{n}_1 \label{g1}
\end{align}
\end{subequations}

The linear cavity Hamiltonian following the displacement transformation takes the form: 
\begin{equation}
\hat{H}^d_2 = -\Delta_2\hat{d}_2^{\dagger}\hat{d}_2 + \widetilde{A}\hat{\sigma}_z(e^{i\beta}\hat{d}_2
+e^{-i\beta}\hat{d}_2^{\dagger} + \frac{1}{\sqrt{\bar{n}_2}}\hat{d}_2^{\dagger}\hat{d}_2 )
\label{H2d}
\end{equation}
where 
\begin{align}
\widetilde{A} = \sqrt{\bar{n}_2}A
\label{dressedA}
\end{align}
is the dressed dispersive qubit-cavity coupling. The linear term in Eq.~(\ref{H2d}) is a qubit state-dependent drive on the linear cavity that encodes qubit state information in the linear cavity quadratures. In contrast, the nonlinear term provides a qubit-state dependent shift of the linear cavity drive detuning; this term would be of no consequence without the linear driving terms. Furthermore, we show later that for a strongly detuned linear cavity and a weak dispersive coupling, this shift in linear cavity drive detuning is a relatively small effect. Therefore, in calculating how the linear cavity quadratures learn about the qubit state, the final nonlinear term can be ignored.

However, the backaction due to this nonlinear term can be significant, since we anticipate the linear cavity quadratures to experience some amplification by virtue of the coupling to the nonlinear cavity. Large fluctuations in linear cavity photon number can contribute to a ``quadratic'' backaction force noise; we characterize this important effect in detail and show that it can be suppressed using appropriate parameter choices, another key advantage of the PARI.

Finally, consider the form of the cavity-cavity coupling Hamiltonian, $\hat{H}_{12}^d$ in the displaced interaction picture. As already mentioned, we take the drive frequencies to satisfy $\omega_{d2} = 2 \omega_{d1}$, allowing us to make a further rotating wave approximation on this coupling (see discussion after Eq.~(\ref{driveCond})). Retaining only the leading (quadratic) terms in the displaced, rotating frame as before, we obtain:
\begin{align}
\hat{H}_{12}^d =  \Big[ &\widetilde{\lambda}_{12}^{(1)}(e^{i\mu_{12}}\hat{d}_1^{\dagger}\hat{d}_2 +
h.c.) \nonumber \\ 
+ &\frac{\widetilde{\lambda}_{12}^{(2)}}{2}(e^{i\mu_{12}}e^{-i2\beta}(\hat{d}_1^{\dagger})^2  + h.c.) \Big],
\label{H12d}
\end{align}
where the dressed intercavity couplings are defined as:
\begin{align}
	\widetilde{\lambda}_{12}^{(j)} = \sqrt{\bar{n}_j}\lambda_{12} \,\,\,\,\,\,\,\,\, (j=1,2)
\label{dressedL12}
\end{align}
and
\begin{equation}
	\mu_{12} = \mu_{2} - 2\mu_{1},
	\label{mu}
\end{equation}
where $\mu_j = \textrm{arg }(\alpha_j)$ is the phase of the cavity-$j$ classical amplitude (cf.~Eq.~(\ref{modes})). Note that this coupling is the only term
in our Hamiltonian with an explicit dependence on these phases.  Also note that this phase difference $\mu_{12}$ is sensitive to the relative phase $\delta$ between the two drives, and hence can be controlled in an experiment. 

The transformed cavity-cavity coupling Hamiltonian in Eq.~(\ref{H12d}) has two types of terms. The first describes simple photon hopping between the two cavities; this term will allow the nonlinear cavity to know about the qubit.  The second term is not a coupling, but effectively renormalizes the cavity-$1$ Hamiltonian.  It describes an effective parametric driving, and thus modifies the strength of the parametric drive term arising from the Kerr nonlinearity in cavity-$1$. We thus define an effective parametric drive strength $\varg_{\mathrm{eff}}$ as:
\begin{equation}
-i\varg_{\mathrm{eff}} = (\varg_1 - \widetilde{\lambda}_{12}^{(2)}e^{i\mu_{12}})e^{-i2\beta} \equiv
|\varg_{\mathrm{eff}}|e^{i(\phi_\varg-2\beta)} 
\label{geff}
\end{equation}
For simplicity, we now make a choice for the reference phase $\beta$, defining $\beta = \phi_{\varg}/2 + \pi/4$, such that
$\varg_{\mathrm{eff}}$ is purely real, $\varg_{\mathrm{eff}} = |\varg_{\mathrm{eff}}|$\footnote{The input noise operators
defined in Eq.~(\ref{drives}) also depend on $\beta$, but the noise correlators are independent of this overall phase and hence it is not necessary to keep track of this.}.

\subsection{Mapping to a single effective nonlinear cavity}
\label{subsec:Mapping}

\subsubsection{Eliminating the linear cavity}

Our driven two-cavity system has been reduced to a linear system of equations, Eqs.~(\ref{heisenbergD}).  For further analysis, it is useful to exactly eliminate cavity-$2$ from the problem, resulting in a description only involving the nonlinear cavity (albeit with modified properties and driving noises). This will be useful for calculating the gain and output noise of our detector; it also directly mirrors our intuition that the linear cavity acts as a tunable coupler between the qubit and the nonlinear cavity.  The elimination procedure is described in Appendix~\ref{appendixEffectiveM}. Introducing canonical quadratures for the displaced cavity fields via $\hat{d}_j = (\hat{x}_j + i \hat{p}_j) / \sqrt{2}$, the Fourier-transformed equations for for the nonlinear cavity take the form:
\begin{align}
	\begin{pmatrix}
		\hat{x}_1[\omega] \\
		\hat{p}_1[\omega]
	\end{pmatrix}
	&=\mathbf{M}_{\mathrm{eff}}^{-1}[\omega] 
	\begin{pmatrix}
		\hat{x}_e^{\mathrm{in}}[\omega] \\
		\hat{p}_e^{\mathrm{in}}[\omega]
	\end{pmatrix}
\label{2DSys}
\end{align}
where the operators $\hat{x}_e^{\rm in}$,~$\hat{p}_e^{\rm in}$ describe both the incident noise on the system as well as the signal associated with the qubit (see Eq.~(\ref{driveEff}) in Appendix~\ref{appendixEffectiveM}). The matrix $\mathbf{M}_{\mathrm{eff}}[\omega]$ is the dressed inverse susceptibility (i.e.~inverse Green function) of the nonlinear cavity.  It has the form
\begin{align}
	\mathbf{M}_{\mathrm{eff}}[\omega]  &\equiv  \mathbf{M}_1[\omega] -i \left(\widetilde{\lambda}_{12}^{(1)}\right)^2 \!\! \mathbf{\Sigma[\omega]}.  
\label{MeffW}
\end{align}
Here, $\mathbf{M}_1$ is the nonlinear cavity inverse susceptibility in the absence of the photon-hopping term in Eq.~(\ref{H12d}); it simply corresponds to a detuned degenerate  parametric amplifier with parametric interaction strength $\varg_{\mathrm{eff}}$ [cf.~Eq.~(\ref{geff})].  In contrast, $\mathbf{\Sigma[\omega]}$ is an effective self-energy matrix, describing the dynamical modification of the nonlinear cavity's properties due to the photon-hopping coupling in Eq.~(\ref{H12d}).  The full form of these matrices is given in Eqs.~(\ref{M1M2}),~(\ref{selfEnergy}) in Appendix~\ref{appendixEffectiveM}.

In the analysis to follow, we will focus on the low-frequency properties of the PARI system.  Focusing on $\omega = 0$ in our interaction picture, we find that the dressed cavity-$1$ susceptibility has the general form expected of a parametric amplifier driven by a detuned pump:
\begin{equation}
\mathbf{M}_{\mathrm{eff}}[0] = 
\begin{bmatrix}
\varg_{\mathrm{eff}} - \frac{\kappa_{\mathrm{eff}}[0]}{2} & -\widetilde{\Delta}_{\mathrm{eff}}[0] \\
\widetilde{\Delta}_{\mathrm{eff}}[0] & -\varg_{\mathrm{eff}} - \frac{\kappa_{\mathrm{eff}}[0]}{2} 
\end{bmatrix}
\label{Meff}
\end{equation}
where the effective nonlinear cavity detuning $\widetilde{\Delta}_{\mathrm{eff}}$ and damping rate $\kappa_{\mathrm{eff}}$ are given by
\begin{align}
	\widetilde{\Delta}_{\mathrm{eff}}[0]&\equiv \widetilde{\Delta}_{\mathrm{eff}} = \widetilde{\Delta}_1 - \left(\widetilde{\lambda}_{12}^{(1)}\right)^2 \!\! \Delta_2|\chi_{2}|^2    \nonumber \\
	\kappa_{\mathrm{eff}}[0] &\equiv \kappa_{\mathrm{eff}}= \kappa_1 + \left(\widetilde{\lambda}_{12}^{(1)}\right)^2 \!\! \kappa_2|\chi_{2}|^2  
\label{paramse}
\end{align}
Note that the general structure of $\mathbf{M}_{\rm eff}$ remains even if $\lambda_{12} = 0$:  a single driven Kerr cavity also has linearized dynamics equivalent to a detuned, degenerate parametric amplifier
\cite{Laflamme2011}.  We have thus mapped our system onto an effective single-cavity system.

It is worth noting that the contributions to $\mathbf{M}_{\rm eff}$ from the self-energy (i.e.~photon hopping) are $\propto \lambda_{12}^2$, 
whereas the direct parametric driving due to cavity-$2$ yields terms that are first order in $\lambda_{12}$ [cf. Eq.~(\ref{geff})].  As a result, it is this second
effect that describes the dominant perturbation of the nonlinear cavity due to the cavity-cavity coupling.

\subsubsection{Amplified and Squeezed Quadratures}

We thus see that similar to a single driven Kerr cavity, the PARI system can be mapped onto an effective detuned degenerate parametric amplifier model.  To make the amplification generated by such a system clearer, it is useful to introduce two new canonical cavity quadratures~\cite{Laflamme2011}. We first introduce the angle $\theta_e$ to parametrize the relative size of the effective pump
detuning to the parametric driving strength [Eqs.~(\ref{paramse}) and (\ref{geff}) respectively],
\begin{equation}
\sin \theta_{e} = \widetilde{\Delta}_{\mathrm{eff}}/\varg_{\mathrm{eff}}.
\label{thetae}
\end{equation}
We will see that having amplification necessarily requires $|\widetilde{\Delta}_{\mathrm{eff}}| < \varg_{\mathrm{eff}}$, hence we can take
$-\pi/2$ $\leq$ $\theta_e$ $\leq$ $\pi/2$ for all regimes of interest.  
In terms of the displaced nonlinear cavity operators $\hat{d}_1,\hat{d}_1^{\dagger}$, the rotated quadratures $\hat{X}_e,\hat{P}_e$ are then defined as
\begin{align}
\hat{X}_e &= \frac{1}{\sqrt{2}} \left( e^{-i\theta_e/2} \hat{d}_1 + e^{i\theta_e/2} \hat{d}^{\dagger}_1 \right) \nonumber \\
\hat{P}_e &= \frac{-i}{\sqrt{2}} \left( e^{-i\theta_e/2} \hat{d}_1 - e^{i\theta_e/2} \hat{d}^{\dagger}_1 \right) 
\label{rotQuads}
\end{align}

By expressing $\left(  \mathbf{M}_{\mathrm{eff}} \right)^{-1}$ in this basis, the solutions of the nonlinear cavity equations of motion (at $\omega=0$) take the form
\begin{subequations}\label{grp}
	\begin{align}
	\hat{X}_e = - &\left[ (\chi_{e-})\hat{X}_e^{\mathrm{in}} - \tan \theta_e
	(\chi_{e-} - \chi_{e+})
	\hat{P}_e^{\mathrm{in}}  \right]  \label{Xe} \\
	&~~~~~~~~~\hat{P}_e = -(\chi_{e+})\hat{P}_e^{\mathrm{in}} \label{Pe}
\end{align}
\end{subequations}
where $\hat{X}_e^{\mathrm{in}}, \hat{P}_e^{\mathrm{in}}$ are the corresponding quadratures of the effective input fields incident on cavity-$1$, and where the 
effective susceptibilities $\chi_{e\pm}$ are given simply by:
\begin{subequations}\label{grp}
\begin{align}
	\chi_{e+}[0] &\equiv \chi_{e+} = \left[ \kappa_{\mathrm{eff}}/2 + \sqrt{\varg_{\mathrm{eff}}^2 - \widetilde{\Delta}_{\mathrm{eff}}^2} \right]^{-1}  \label{chiEP} \\
	\chi_{e-}[0] &\equiv \chi_{e-} = \left[ \kappa_{\mathrm{eff}}/2 - \sqrt{\varg_{\mathrm{eff}}^2 - \widetilde{\Delta}_{\mathrm{eff}}^2} \right]^{-1} 
\label{chiEM} 
\end{align}
\end{subequations}

Amplification in a detuned DPA model emerges when one tunes parameters so that the susceptibility $\chi_{e-}$ diverges.  This results both in extremely large fluctuations of the quadrature $\hat{X}_e$, and in 
amplification of signals driving this quadrature.  The standard parametric photon number gain $\mathcal{G}$ is derived from the system scattering matrix (see e.g.~\cite{Laflamme2011}):
\begin{equation}
	\mathcal{G}[\omega] = \left| 1 - \kappa_1 \chi_{e-}[\omega] \right|^2 \overset{}{\underset{\chi_{e-}\to
	\infty}{\longrightarrow}} \kappa_1^2\chi_{e-}^2
\label{parGain}
\end{equation}
We thus see that $\hat{X}_e$ represents the amplified quadrature of the cavity.  In contrast, the orthogonal quadrature $\hat{P}_e$ is {\it not} the squeezed quadrature of the cavity:  signals incident in this quadrature can also emerge in $\hat{X}_e^{\rm out}$ with amplification.  As discussed in Ref.~\onlinecite{Laflamme2011}, the finite effective detuning $\tilde{\Delta}_{\rm eff}$ leads to the fact that standard amplified and squeezed quadratures are not orthogonal.

For a single driven Kerr cavity, one can achieve a diverging $\chi_{e-}$ by driving the system close to a point of bifurcation.  In the PARI system, we have seen that the basic classical equations determining the classical amplitudes are modified; we thus first need to understand whether a similar bifurcation (and diverging $\chi_{e-}$) can still be attained, at cavity-cavity coupling strengths $\lambda_{12}$ sufficient to allow efficient qubit measurements.

\subsection{Bifurcation for weak cavity-cavity couplings} 
\label{weakCoupling}

To ensure the PARI also exhibits a simple bifurcation in its classical dynamics (and hence amplification), we will focus on regimes of weak cavity-cavity coupling $\lambda_{12}$, such that this only weakly perturbs the bifurcation that would occur in the uncoupled nonlinear cavity. Using equation Eq.~(\ref{classicalA1}) for the classical cavity-$1$ amplitude, and the modification of the parametric driving strength in Eq.~(\ref{geff}), we see that this requires:
\bse
\label{eqs:CouplingBounds}
\begin{eqnarray}
	\lambda_{12}\sqrt{\bar{n}_{2}} & \ll &  2\Lambda \bar{n}_1  
	\label{weakDC} \\
	\lambda_{12} & \ll & 2\sqrt{\Lambda/|\chi_2|} \label{weakDC2}
\end{eqnarray}
\ese
where $\bar{n}_{2}$ is the linear cavity photon number. The first condition ensures that the coupling-induced parametric driving term in Eq.~(\ref{classicalA1}) ($\propto \eta$) is much smaller than the main Kerr term ($\propto \Lambda$).  Note that this condition needs to be satisfied self-consistently, as $\bar{n}_1$ follows from solutions to the classical equations of motion, and thus depends weakly on $\lambda_{12}$. The second equation, and the requirement of a strongly detuned linear cavity $\Delta_2 \gg \kappa_2$, ensures that the imaginary Kerr constant induced by the cavity-cavity coupling (cf.~Eq.~(\ref{complexKerr})) is also much weaker than the main Kerr term $\Lambda$. These conditions together define the weak coupling regime in terms of the strength of the cavity-cavity coupling $\lambda_{12}$, where we expect simple bifurcation physics to be retained.

\begin{figure}[t]
\includegraphics[scale = 0.62]{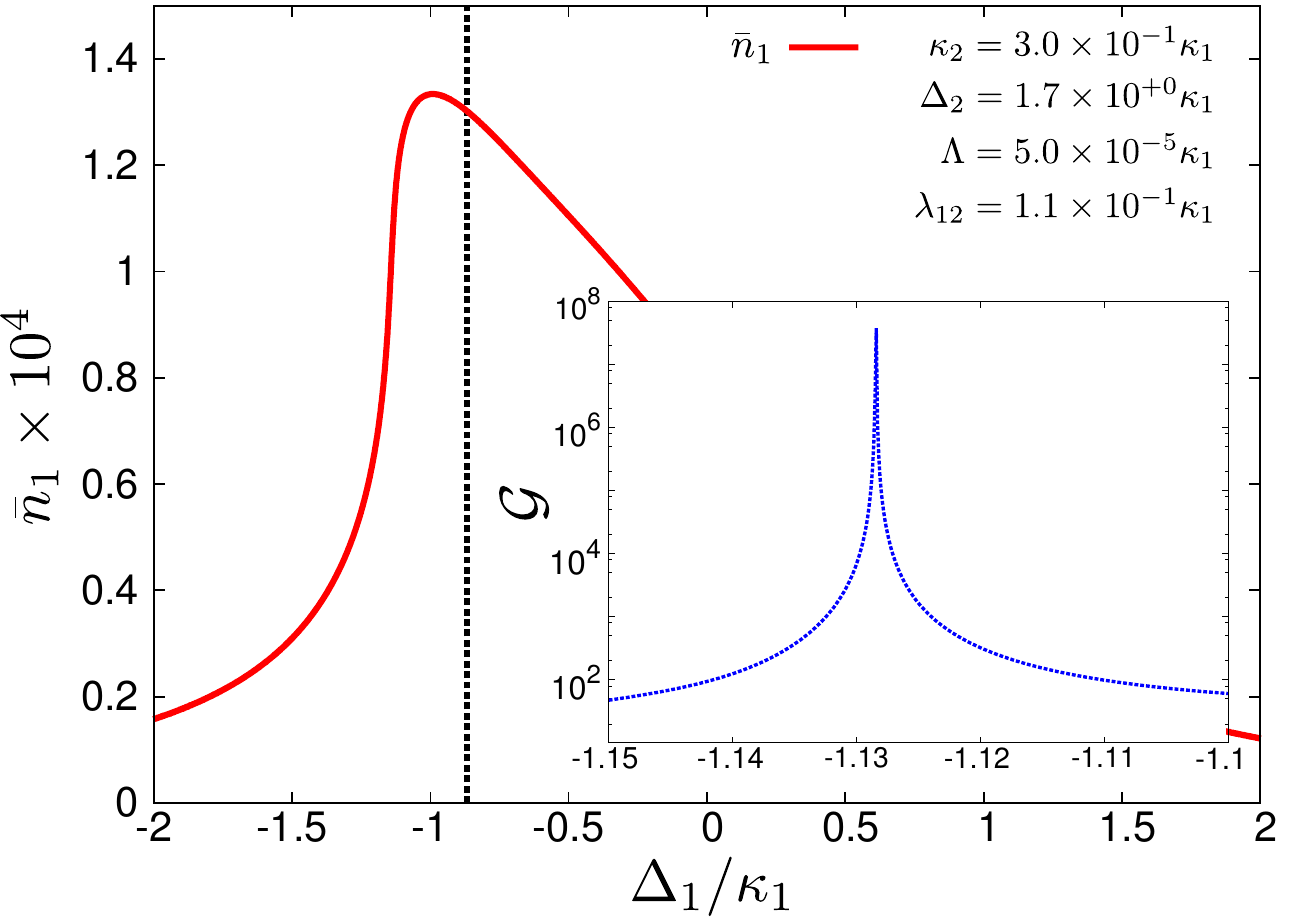}
\caption{Plot of the nonlinear cavity photon number $\bar{n}_1$, against nonlinear cavity drive detuning $\Delta_1$.  We take a cavity-cavity coupling $\lambda_{12}$ weak enough (as per Eqs.~(\ref{eqs:CouplingBounds})), to ensure a small perturbation of bifurcation physics; parameter values are listed on the plot.  The detuning at which the bifurcation occurs when $\lambda_{12} = 0$ is indicated by the dashed vertical line. The nonlinear cavity drive is adjusted so that the parametric gain $\mathcal{G}$, is large, thereby pushing the nonlinear cavity close to, but not beyond the bifurcation,. The inset plot indicates $\mathcal{G}$ versus $\Delta_1$; the large maximum value indicates the preservation of amplification behaviour of the nonlinear cavity. }
\label{n1Profiles}
\end{figure}

To demonstrate that these conditions are sufficient, we solve Eq.~(\ref{classicalA1}) using an appropriately weak $\lambda_{12}$, and using realistic values for $\kappa_1$, $\kappa_2$; for reasons that will become clear in section~\ref{optimalParameters}, we require $\Delta_2 \sim \kappa_1 \gg \kappa_2$. Shown in Fig.~\ref{n1Profiles} are resulting plots of the classical cavity-$1$ intensity $\bar{n}_1$ versus the cavity-$1$ drive detuning $\Delta_1$. The steepness of the plotted response curves indicates proximity to a bifurcation, and the possibility of large parametric gain.
This is confirmed in the inset, where  we plot the corresponding parametric photon number gain $\mathcal{G}[\omega=0]$ for the modified nonlinear cavity.

The upshot of the analysis here is that as long as $\lambda_{12}$ is weak enough to satisfy Eqs.~(\ref{eqs:CouplingBounds}), one can still have simple bifurcation in the classical dynamics, and hence large parametric gain.
We do see modifications in the position of the bifurcation point and the cavity-$1$ drive strength needed to reach this point; further details of these modifications are given in Appendix~\ref{appendixClassical}, including a perturbation theory approach to determining how the bifurcation shifts.  

While a weak $\lambda_{12}$ ensures simple bifurcation physics, the coupling 
also needs to be sufficiently strong that the qubit signal in the nonlinear cavity output is suitably large.  In Appendix~\ref{appendixMinL12}, we calculate a corresponding minimum useful value of $\lambda_{12}$.  Below this value, the coupling to the nonlinear cavity is so weak that it would be more efficient to use the linear cavity output field to read out the qubit state.  We find
\begin{equation}
	\lambda_{12}^{\mathrm{min}} \sim \sqrt{ \frac{\kappa_1\kappa_2}{ \bar{n}_1 \mathcal{G}}}
	\label{l12Min}
\end{equation}
For sufficiently large parametric $\mathcal{G}$, this minimum value is very far from the upper limits established by Eqs.~(\ref{eqs:CouplingBounds}).  Thus, there is a large range of cavity-cavity coupling strengths where one can retain the simple bifurcation physics of a single nonlinear cavity, while still having enough qubit information transferred to the nonlinear cavity.


\section{Analysis and optimization of detector noise and response}
\label{sec:results}

In the previous section, we demonstrated that the PARI system exhibits a bifurcation similar to a single, driven Kerr cavity, and that  near such a point, it can be mapped on to a degenerate parametric amplifier (driven by a detuned pump). We now reintroduce the coupling to the qubit, and analyze the properties of the system as a continuous qubit detector. The PARI system has a large number of parameters, which we summarize in Table~\ref{availableParams}.  Thus, a key issue to address is how to pick these parameters to ensure optimal system performance.  

We will start our analysis by addressing what seems to be the largest flaw in the design:  the qubit is not sufficiently protected from the large noise associated with the driven nonlinear cavity, and hence will experience backaction noise well in excess of the quantum limit.  We will show that this expectation is false:  with appropriate parameter tuning, the qubit can be made immune to the large fluctuations in the nonlinear cavity.  We will then turn to analyzing the output characteristics of the detector:  both the gain (i.e. the size of the qubit signal) and the measurement rate (i.e. the size of this signal relative to the amount of noise in
the detector output).

\begin{table}[t]
\begin{center}
\begin{tabular}{c|c}
\hline
\multicolumn{2}{c}{Conditions to ensure a simple bifurcation} \\
\hline 
\rule{0pt}{3ex} Weak coupling regime & $ \sqrt{\frac{\kappa_1\kappa_2}{\bar{n}_1\mathcal{G} }} \ll \lambda_{12} \ll
\frac{2\Lambda\bar{n}_1}{\sqrt{\bar{n}_{2}}} <  2\sqrt{\frac{\Lambda}{|\chi_2|}}  $  \\
Strongly detuned cavity-$2$  & $\Delta_2 \gg \kappa_2$ \\
\hline
\multicolumn{2}{c}{Parameter list} \\
\hline
\multicolumn{2}{c}{Cavity $j$: damping rate $\kappa_j$, drive strength $\bar{a}_j^{\rm in}$, drive detuning $\Delta_j$} \\
\multicolumn{2}{c}{Nonlinear cavity Kerr constant, $\Lambda$} \\
\multicolumn{2}{c}{Intercavity coupling, $\lambda_{12}$} \\
\multicolumn{2}{c}{Relative drive phase, $\delta$} \\
\multicolumn{2}{c}{Dispersive coupling, $A$} \\
\hline
\end{tabular}
\caption{Conditions defining the parameter regimes that must be adhered to for the nonlinear cavity to be modified only weakly by its coupling to the linear cavity - the `weak coupling' regime. The lower table lists the large number of tunable parameters in the PARI scheme that help optimize performance.}
\label{availableParams}
\end{center}
\end{table}

\subsection{Backaction noise}

The basic dispersive coupling between the qubit and cavity-$2$ was given in Eq.~(\ref{Hsigmaz}), where we introduced the backaction force operator $\hat{F} = A \hat{a}^\dagger_2 \hat{a}_2$.  For a weak qubit-detector coupling $A$, the low frequency fluctuations $\hat{F}$ will dephase an initial superposition in the qubit.  This dephasing is described by the rate (see, e.g., \cite{Blais04,ClerkRMP})
\begin{equation}
	\Gamma_{\varphi} =2 S_{FF}[0], 
\label{gammaPhi}
\end{equation}
where $S_{FF}[\omega]$ is the noise spectral density of $\hat{F}$, evaluated at zero qubit coupling:
\begin{equation}	
	S_{FF}[\omega] \equiv \int_{-\infty}^{\infty} dt~e^{i \omega t} \langle \hat{F}(t) \hat{F}(0) \rangle_0.
\label{SFFdef}
\end{equation}

Working in the displaced interaction picture [cf.~Eq.~(\ref{modes})], the backaction force operator takes the form:
\begin{eqnarray}
	\hat{F} & = & A\sqrt{\bar{n}_2}(e^{i\beta}\hat{d}_2 + h.c.) + A \hat{d}^\dagger_2 \hat{d_2} \equiv \hat{F}_{\rm L} + \hat{F}_{\rm Q} 
\label{Fop}
\end{eqnarray}
We have split the backaction force operator into parts that are linear and quadratic in the $\hat{d}_2$ operators. For a suitably strong cavity-$2$ drive, $\bar{n}_2 \gg 1$, the dominant noise contribution will come from $\hat{F}_L$; we thus first study this quantity.

\subsubsection{Linearized backaction and backaction avoidance}

Using the linearized Heisenberg-Langevin equations for our system, we can decompose the Fourier-transformed Heisenberg-picture operator $\hat{F_{\rm L}}[\omega]$ as 
\begin{equation}
	\hat{F}_{\rm L}[\omega] =   \hat{F}_{{\rm L}}^{\rm int}[\omega] + 
		\widetilde{\lambda}_{12}^{(1)} \left(  f_X[\omega] \hat{X}_e[\omega] + f_P[\omega] \hat{P}_e[\omega] \right);
\label{FE}
\end{equation}
see Eq.~(\ref{fINT}) in Appendix~\ref{appendixF} for details.  Here, $\hat{F}_{{\rm L}}^{\rm int}$ is the $\lambda_{12}$-\emph{independent} contribution to $\hat{F}_{\rm L}$.  It describes vacuum noise incident from the cavity-$2$ input port which has not entered cavity-$1$, and hence has not been amplified by the nonlinear cavity.  The remaining terms describe noise contributions emanating from the nonlinear cavity. We have written them in terms of the quadratures of the nonlinear cavity $\hat{X}_e$
and $\hat{P}_e$, as introduced in Sec.~\ref{subsec:Mapping} (cf.~Eq.~(\ref{Xe}) and (\ref{Pe})).  $f_X[\omega], f_P[\omega]$ are real coefficients which are independent of the cavity-$1$ susceptibility (cf.~Eq.~(\ref{fINT})).  

As discussed in Sec.~\ref{subsec:Mapping}, it is only the $\hat{X}_e$ quadrature which experiences amplification due to the effective parametric interaction; thus, this is the only term in Eq.~(\ref{FE}) which describes a diverging, amplified noise contribution near the bifurcation.  Eliminating this large noise contribution thus involves tuning parameters such that the coefficient $f_X[\omega]=0$ over frequencies of interest. We have in general:
\begin{align}
	f_X[\omega] = -\frac{\widetilde{A}}{\sqrt{2}}\left[\chi_{2}^*[-\omega]e^{i(\mu_{12} - \nu[\omega])} + \chi_2[\omega]e^{-i(\mu_{12} - \nu[\omega])}\right] 
\label{FXFull} 
\end{align}
Here, the phase $\mu_{12}$ is the relative phase between the classical cavity amplitudes (cf.~Eq.~(\ref{mu})); it determines the effective phase of the linearized cavity-cavity coupling (cf.~Eq.~(\ref{H12d})). The remaining phase angle $\nu[\omega]$ is
\begin{equation}
	\nu[\omega] = \phi_{\varg}/2 + \theta_e[\omega]/2 + 3\pi/4,
\label{nu}
\end{equation}
where $\phi_{\varg}$ is the phase of the effective parametric interaction in Eq.~(\ref{geff}), and $\theta_e[\omega]$ characterises the rotation which defines the amplified $\hat{X}_e$ quadrature at each frequency (cf.~Eq.~(\ref{thetae})). For reasons that will become clear shortly, we focus on the $\omega = 0$ regime. Writing the bare cavity-$2$ susceptibility defined in Eq.~(\ref{chi2}) as $\chi_2[\omega] = |\chi_2[\omega] | \exp( i \phi_2[\omega])$, $f_X[\omega=0]$ takes the form
\begin{align}
	f_X[0] &= -\sqrt{2}\widetilde{A}~|\chi_{2}[0]|~\cos (\mu_{12} -\phi_2[0] - \nu[0]). 
\label{FX} 
\end{align}
Therefore, at zero frequency,  one can make $f_X[0]$ (and hence the largest noise contribution to $\hat{F}_{\rm L}[\omega]$) vanish by an appropriate choice of the phase $\mu_{12}$; this phase can in turn be varied by tuning the relative phase $\delta$ between the two cavity drives.  One needs:
\begin{equation}
\mu_{12} - \phi_2[0] - \nu[0] = (2M+ 1)\frac{\pi}{2},~~~M \in \mathbb{Z}
\label{reducedF}
\end{equation}

In this zero-frequency case, when this \emph{reduced backaction condition} is satisfied, the backaction noise in $\hat{F}_L$ is 
\begin{align}
	S_{F_{\rm L}F_{\rm L}}[0] 
	& \simeq S_{F_{\rm L}F_{\rm L}}^{\rm int}[0]  +  S_{F_{\rm L}F_{\rm L}}^{\rm NL}[0], 
	\label{Sff} \\
	S_{F_{\rm L}F_{\rm L}}^{\rm int}[0] &= \widetilde{A}^2   \kappa_2 |\chi_2|^2\\ 
	S_{F_{\rm L}F_{\rm L}}^{\rm NL}[0] &=\widetilde{A}^2 \! \left(\widetilde{\lambda}_{12}^{(1)}\right)^2 \!\! |\chi_2|^2 \kappa_1\chi_{e+}^2 + \ldots ,
	\label{SFLFL}
\end{align}
see Appendix~\ref{appendixF} for an outline of the full calculation. $S_{F_{\rm L}F_{\rm L}}^{\rm int}[0]$  describes the unamplified contribution of vacuum noise incident on cavity-$2$, while $S_{F_{\rm L}F_{\rm L}}^{\rm NL}[0]$ describes the contribution of all noise processes involving the nonlinear cavity (including correlations between the two terms in Eq.~(\ref{FE})).  This latter contribution is solely due to the non-amplified cavity quadrature (it is just a function of non-diverging susceptibility $\chi_{e+}$, which scales as $1/\kappa_1$ near the bifurcation),  and thus {\it the backaction noise and dephasing is independent of the amplified vacuum noise fluctuations in the nonlinear cavity}.  We see that careful tuning of the drive phases gives the system a kind of directionality: the nonlinear cavity will amplify the qubit signal, but the qubit will not see amplified vacuum noise emanating from the nonlinear cavity. This protection against backaction is impossible in systems where a qubit is directly, dispersively coupled to a nonlinear cavity, as there is no way to tune the cavity quadrature seen by the qubit; this is why these systems have a backaction far in excess of the quantum limit \cite{Laflamme2011}.

We stress that with appropriate parameter choices,  enforcing Eq.~(\ref{reducedF}) at $\omega = 0$ will be enough to guarantee a noise suppression over a wide bandwidth.  To see this, note that all the frequency dependence in this equation is ultimately related to the frequency dependence of $\chi_2[\omega]$, the bare cavity-$2$ susceptibility (cf.~Eq.~(\ref{chi2})).  By choosing a cavity-$2$ drive tone that is far detuned from resonance (i.e.~$\Delta_2 \gg \kappa_2$), $\chi_2[\omega]$ will effectively become frequency independent over the amplification bandwidth.  As such, ensuring Eq.~\ref{reducedF} holds at zero frequency ensures that it holds over this entire bandwidth. Fig.~\ref{GainBandwidth} in Appendix~\ref{appendixGain} indicates this frequency dependent behaviour and confirms that the zero frequency approximation is valid.

\subsubsection{Quadratic backaction}

We now turn to the fluctuations of $\hat{F}_{\rm Q}$, the second term in Eq.~(\ref{Fop}) which is proportional to the photon number $\hat{d}^\dag_2 \hat{d}_2$.  While this term is smaller by a factor of $\bar{n}_2$ than the leading contribution, the fact that it is quadratic in field operators implies that it can still contribute significant noise near bifurcation. Furthermore, calculating the autocorrelation of $\hat{d}_2^{\dagger}\hat{d}_2$ involves a convolution of $\hat{d}_2,\hat{d}_2^{\dagger}$ correlators, which incorporates noise at all frequencies. The calculation of the various noise components of this term is outlined in section~\ref{appendixF}; in the relevant large gain limit, the dominant contributions to the quadratic backaction can be approximately written as:
\begin{equation}
	S_{F_{\rm Q} F_{\rm Q}}[0] \simeq   A^2\left(\widetilde{\lambda}^{(1)}_{12}\right)^4 \!\!\!\! \sum_{\omega=0,~\Delta_2} \!\! \frac{\mathcal{G}[\omega]^2}{\kappa_1^2} \cdot|\chi_2[\omega]|^4\cdot \Omega[\omega],
\label{quadraticBA}
\end{equation}
where $\mathcal{G}[\omega]$ (parametric photon number gain) and$|\chi_2[\omega]|$ (cavity-$2$ susceptibility) parametrize the nonlinear and linear cavity responses respectively, and $\Omega[\omega]$ is an effective bandwidth. Near $\omega = 0$ and $\omega = \Delta_2$, and over a frequency range specified by the bandwidth $\Omega[\omega]$, the nonlinear and linear cavity responses respectively are strongest, indicating a large contribution to the backaction noise.  The $\mathcal{G}^2$ dependence here 
would also occur in a single nonlinear cavity directly coupled to a qubit.  While there is no simple tuning possible in the PARI that lets one completely suppress the amplified noise contribution to $\hat{F}_{\rm Q}$, one can strongly reduce the relative importance  of $\hat{F}_Q$ compared to other terms by increasing the cavity-$2$ photon number:  $S_{F_{\rm Q}F_{\rm Q}}$ depends on the bare dispersive coupling $A$ and \emph{not} the dressed $\widetilde{A}$ (cf. Eq.~(\ref{dressedA})). In the limit where $\bar{n}_2 \to \infty$ while the dressed dispersive coupling $\widetilde{A}$ is kept fixed, the quadratic backaction formally goes to zero.  The PARI system also has additional tunability compared to a single-cavity system which
allows a further suppression of this term;  for a full discussion, see Appendix~\ref{appendixF}.
%
%

\subsection{Detector response and amplifier forward gain}

Having characterized the backaction noise of the PARI detector, we now examine how information on the qubit state appears at the detector output.  Referring back to Fig.~\ref{boxSchematic}, recall that information about the qubit signal is obtained by measuring the field {\it  leaving the nonlinear} cavity; this signal will thus benefit from amplification in this cavity.  We consider a homodyne measurement, where one measures the quadrature of this output field containing the maximal amount of information on the qubit state.  Details on the basics of homodyne measurement and their description can be found in, e.g., Refs.~\cite{ClerkRMP, milburn2010} 

We first want to understand the signal produced by the qubit, i.e. the average homodyne current associated with the two qubit $\sigma_z$ eigenstates.  Note from  Eq.~(\ref{H2d}) that in our displaced frame, the qubit both acts as an effective linear driving force on cavity-$2$, as well as shifts its frequency.  We focus initially on the first effect, which is enhanced by the average cavity-$2$ photon number $\bar{n}_2$. Using the effective nonlinear cavity picture developed in Sec.~\ref{subsec:Mapping}, this qubit-induced drive will act as a driving force on both the $\hat{X}_e$ and $\hat{P}_e$ nonlinear cavity quadratures; it thus yields new terms in the driving fields in Eqs.~(\ref{Xe}),~(\ref{Pe}):  
\begin{align}
\hat{X}_e^{\mathrm{in}} \to \hat{X}_e^{\mathrm{in}} - \left( \sqrt{
\frac{2}{\kappa_{\mathrm{eff}}}}~\widetilde{A}~\widetilde{\lambda}_{12}^{(1)}~|\chi_2|\mathrm{Im} \left[e^{i(\mu_{12} +\phi_2 - \nu)}
\right] \right)\hat{\sigma}_z \nonumber \\
\hat{P}_e^{\mathrm{in}} \to \hat{P}_e^{\mathrm{in}} + \left( \sqrt{
\frac{2}{\kappa_{\mathrm{eff}}}}~\widetilde{A}~\widetilde{\lambda}_{12}^{(1)}~|\chi_2|\mathrm{Re} \left[e^{i(\mu_{12} + \phi_2 - \nu)} \right] \right) \hat{\sigma}_z
\label{qubitSub}
\end{align}
The measured homodyne current $\hat{I}_1$ is just proportional to a quadrature of the output field leaving cavity-$1$,  i.e.~$\hat{I}_1 \propto \left( e^{i \phi_h} \hat{d}_{1,{\rm out}} + h.c. \right)$.  For simplicity, we assume in what follows an optimal choice of quadrature, by appropriate choice of the homodyne angle $\phi_h$. Using standard input-output theory, we can easily calculate its stationary, average value. Considering the case where we are near bifurcation (and hence have a large parametric photon number gain $\mathcal{G}$, cf.~Eq.~(\ref{parGain})), we find to leading order
\begin{eqnarray}
	\langle \hat{I}_1 \rangle & \equiv & \chi_{IF}[0] \langle \hat{\sigma}_z \rangle, \nonumber \\
	 \chi_{IF}[0] & \simeq &
	 \sqrt{2}~\widetilde{A}~\widetilde{\lambda}_{12}^{(1)}~|\chi_{2}|~\sqrt{\mathcal{G}}~\rho_e,
	\label{gain}
\end{eqnarray}
where $\chi_{IF}[0]$ is the zero-frequency detector response coefficient or ``forward gain" \cite{ClerkRMP}, and the angle-dependent prefactor $\rho_e$ is given by
\begin{equation}
	\rho_e = \sec \theta_e \sin(\mu_{12} + \phi_2 -\nu + \theta_e).
	\label{rhoE}
\end{equation}
All quantities are at zero frequency. For details of the full calculation, see Appendix~{\ref{appendixGain}.

As expected, the response of the detector is enhanced by the large parametric gain $\mathcal{G} \gg 1$ generated when the PARI system is operated near bifurcation.  Further, note that tuning the system
to satisfy the reduced backaction condition [cf.~Eq.~(\ref{reducedF})] {\it does not make the forward gain vanish}.  The reduced backaction condition ensures that the qubit is coupled to the non-amplified quadrature $\hat{P}_e$ of the nonlinear cavity; it thus efficiently drives the orthogonal amplified quadrature $\hat{X}_e$, leading to the large, amplified signal.

We have thus far ignored the second effect of the qubit on the linear cavity, namely a dispersive shift of the linear cavity drive detuning $\Delta_2$ [cf. Eq.~(\ref{H12d})]. It turns out that for the case of a strongly detuned linear cavity and weak dispersive coupling so that $A \ll \Delta_2$ (parameters indicated in Table~\ref{parChoice}), $\Delta_2 \pm A$ produces less than a $1~\%$ change in $\Delta_2$. The resulting effect on the gain is small, and so the effect of the nonlinear term in the qubit-detector coupling can be neglected.

\subsection{Measurement imprecision noise and measurement rate}

We now turn to analyzing the intrinsic noise in the output homodyne current, $\bar{S}_{II}[\omega]$, in the PARI setup; along with the forward gain, this will determine the imprecision of our measurement and the measurement rate (the rate at which information on the qubit state is acquired). The calculation of this intrinsic zero-frequency noise $S_{II}[0]$ (i.e.~calculated in the absence of any qubit coupling) follows straightforwardly from the linearized Heisenberg-Langevin equations for our system; details are presented in Appendix~\ref{appendixGain}. In the large gain limit of interest, we find to leading order
\begin{equation}
	S_{II}[0] \simeq \frac{1}{2} \kappa_{\mathrm{eff}}  \sec^2\theta_e~\mathcal{G},
\label{SII}
\end{equation}
where $\kappa_{\mathrm{eff}}$ is the effective nonlinear cavity damping rate defined in Eq.~(\ref{paramse}).  Apart from this renormalized value of $\kappa_1$ (which is due to the cavity-cavity interaction), this result has the same general form and dependence on $\mathcal{G}$ as for a single Kerr cavity driven near bifurcation~\cite{Laflamme2011}. 

We can now calculate the measurement time $\tau_{\rm meas}$ associated with our setup:  how long will it take to resolve the qubit signal above the noise in the detector output.  For the weak qubit-coupling, linear-response regime we consider, $\tau_{\rm meas}$ has the standard definition \cite{ClerkRMP}
\begin{equation}
	\frac{1}{\tau_{\rm meas}} = \Gamma_{\mathrm{meas}} = \frac{|\chi_{IF}[0]|^2}{2S_{II}[0]} \equiv \frac{1}{2} S_{zz}^{-1}
\label{eq:GammaMeas}
\end{equation}
where $S_{II}[0]$ is the intrinsic zero-frequency noise in the measured homodyne current (i.e. in the absence of any coupling to the qubit).  In the last inequality, we have introduced the imprecision noise spectral density $S_{zz}[0]$, which is just the output noise referred back to the qubit.  

In the large gain limit, making use of Eqs.~(\ref{gain}) and (\ref{SII}), we find
\begin{equation}
	\Gamma_{\mathrm{meas}}  \simeq
		\left[  \left(\widetilde{\lambda}_{12}^{(1)}\right)^2 \!\! |\chi_{2}|^2 \right] \times  \frac{2 \widetilde{A}^2}{\kappa_{\mathrm{eff}}}\sin^2(\mu_{12} + \phi_2 -\nu +  \theta_e)
	\label{meas}
\end{equation}
The nonlinear cavity amplifies both the signal and the noise in the cavity output; as a result, the rate of information gain, $\Gamma_{\mathrm{meas}}$, is independent of $\mathcal{G}$. This is identical to the case where one directly couples a qubit to a nonlinear cavity~\cite{Laflamme2011}. However, unlike this direct coupling case, there is additional tunability in the PARI setup.  In particular, the coupling prefactors in Eq.~(\ref{meas}) (i.e~factor in square brackets) can be tuned to be larger than unity, even if $\lambda_{12}$ is so weak that it does not perturb the bifurcation physics of the nonlinear cavity.

In addition to having a large measurement rate, we also want the sheer magnitude of the output signal to be large enough that the noise of any following amplification stages is irrelevant.  It is useful to quantify the size of the output noise via an effective number of thermal quanta $\bar{n}_{II}$:  if the cavity were linear and in thermal equilibrium, how hot would it have to be to produce the same amount of output noise?  This leads to the definition $S_{II}[0] = \frac{\kappa_1}{2}(1 + 2\bar{n}_{II})$. As both the forward gain $\left|\chi_{IF}[0]\right|^2$ and output noise $S_{II}[0]$ scale as $\mathcal{G}$, we are in good shape: $\bar{n}_{II} \propto \mathcal{G}$, and hence, in principle the magnitude of the output noise and signal can be made large enough to make the noise added in following amplification stages insignificant.




\section{Parameter optimization for large gain, quantum-limited performance}

The previous section suggested that the PARI design should allow one to both produce a large-magnitude qubit signal, as well as a small backaction disturbance of the qubit (one that is immune to the amplified fluctuations of the nonlinear cavity).  In this section, we show this indeed possible with realistic parameter choices, and also explain the rationale governing these choices.

We start by recalling that the fundamental Heisenberg constraint on the noise of any linear detector implies that there is a minimum possible value for the detector backaction: the backaction dephasing cannot be any smaller than the measurement rate \cite{ClerkRMP}. We thus define the efficiency ratio $\eta_{\rm meas}$ which characterizes the size of the actual detector backaction compared to this ideal minimum:
\begin{equation}
	\frac{1}{\eta_{\rm meas}} = \Gamma_{\varphi}/\Gamma_{\mathrm{meas}} = 4S_{zz}S_{FF} \geq 1
\label{SzzSff}
\end{equation}
The quantum limit corresponds to achieving $\eta_{\rm meas}=1$.

\subsection{Procedure for selecting optimal parameters}
\label{optimalParameters}

In choosing parameters, the first step is to take the relative phase of the cavity drives to fulfill the reduced backaction condition of Eq.~(\ref{reducedF}), which eliminates the contribution of amplified noise to the linear backaction operator $\hat{F}_L$.  Once this is done, the remaining contributions to the backaction noise come from Eq.~(\ref{Sff}) (the remaining fluctuations in $\hat{F}_{\rm L}$) and Eq.(\ref{quadraticBA}) (fluctuations in the quadratic backaction operator $\hat{F}_{\rm Q}$); we begin by analysing the former.

The non-amplified fluctuations in $\hat{F}_{\rm L}$ arise from the intrinsic linear cavity backaction and the residual nonlinear cavity backaction. For signal extraction from the nonlinear cavity, the linear cavity backaction represents excess noise 
not directly tied to any information gain. Ideally, we require the residual nonlinear cavity backaction to be larger than this intrinsic linear cavity noise, so that more information is being extracted from the nonlinear cavity than is leaking out the linear
cavity drive port;  from Eq.~(\ref{SFLFL}), this occurs when the dressed coupling $\widetilde{\lambda}_{12}^{(1)}$ satisfies the constraint:
\begin{align}
	\left(\widetilde{\lambda}_{12}^{(1)}\right)^2 > \kappa_1\kappa_2
	\label{l12Constraint}
\end{align}
As $\bar{n}_1$ is itself dependent on $\lambda_{12}$, this is a self-consistent equation for $\lambda_{12}$. As we work near bifurcation, $\bar{n}_1$ will be large and scaling roughly as $\kappa_1/(2\sqrt{3}~\Lambda)$ \cite{Laflamme2011}. For given values of $\kappa_1, \kappa_2$ and $\Lambda$, Eq.~(\ref{l12Constraint}) constrains the minimum value of $\lambda_{12}$.  This then seems to imply that we should couple the cavities as strongly as possible, thus making $\widetilde{\lambda}_{12}^{(1)}$ large. However, an analysis of the quadratic backaction (cf. Eq.~(\ref{quadraticBA})) immediately reveals a problem: the dominant contributions to $S_{F_{\rm Q}F_{\rm Q}}$ depend \emph{quartically} on the same dressed coupling, and therefore are enhanced more strongly than $S_{F_{\rm L}F_{\rm L}}^{\rm NL}$ when $\widetilde{\lambda}_{12}^{(1)}$ is increased.

\begin{table}[t]
\begin{center}
\begin{tabular}{c|c}
\hline
Parameter & Value (MHz) \\
\hline
Nonlinear cavity damping, $\kappa_1$ & $100$ \\
Linear cavity damping, $\kappa_2$, & $30$ \\
Linear cavity detuning, $\Delta_2$ & $170$ \\
Intercavity coupling, $\lambda_{12}$ & $1.10$ \\
Dispersive coupling, $A$ & $1$\\
Kerr constant, $\Lambda$ & $0.005$ \\
Nonlinear cavity detuning, $\Delta_1$ & (see caption) \\
Cavity drive strengths, $\bar{a}_j^{\rm in}$ & (see caption) \\
Relative drive phase, $\delta$ & (see caption) \\
\hline
\multicolumn{2}{c}{Quantum limited performance constraint} \\
\hline
\multicolumn{2}{c}{ \rule{0pt}{3.1ex} $\frac{\kappa_1^2}{\sqrt{\mathcal{G}}} < \left(\widetilde{\lambda}_{12}^{(1)}\right)^2 < \frac{\bar{n}_2(\Delta_2)^2}{\mathcal{G}^{3/2}}$}\\
\multicolumn{2}{c}{ $0.10 < 1.14 < 4.93$} \\
\hline
\multicolumn{2}{c}{Weak coupling constraints} \\
\hline
\multicolumn{2}{c}{\rule{0pt}{3ex} $ \widetilde{\lambda}_{12}^{(2)} < 2\Lambda\bar{n}_1~,~2\sqrt{ \frac{\Lambda\bar{n}_2}{|\chi_2|}} $} \\
\multicolumn{2}{c}{ $0.45 < 0.80~,~0.74$} \\
\hline
\end{tabular}
\caption{Table indicating realistic parameter choices used in Figs.~\ref{GainVsDelta1},~\ref{MeasVsDelta1}, which satisfy the constraints in Eqs.~(\ref{finalConstraint}),~(\ref{L12n2Constraint}) as indicated by the lower tables. The cavity drive strengths $\bar{a}_j^{\rm in}$, relative drive phase $\delta$, and the nonlinear cavity detuning $\Delta_1$ form a parameter space where optimal operating points are found via a numerical optimization routine.}
\label{parChoice}
\end{center}
\end{table}

Fortunately, the PARI system provides enough flexibility of parameter choices to overcome this problem. We enforce self-consistently that both large terms in the quadratic backaction $S_{F_{\rm Q}F_{\rm Q}}$, near $\omega = 0,~\Delta_2$, are smaller than the contribution to $S_{F_{\rm L}F_{\rm L}}$ from the nonlinear cavity. The form of each term is characterised in Appendix~\ref{appendixF}; we find that to suppress the quadratic backaction, the following conditions must hold:
\begin{align}
&S_{F_{\rm L}F_{\rm L}}^{\rm NL} > S_{F_{\rm Q}F_{\rm Q}}^{(1)} \implies \left(\widetilde{\lambda}_{12}^{(1)}\right)^2 < \frac{\bar{n}_2 (\Delta_2)^2}{\mathcal{G}^{3/2}} \nonumber \\
&S_{F_{\rm L}F_{\rm L}}^{\rm NL} > S_{F_{\rm Q}F_{\rm Q}}^{(2)} \implies \left(\widetilde{\lambda}_{12}^{(1)}\right)^2 < \bar{n}_2 (\Delta_2)^2 \! \left(\frac{\kappa_2}{\kappa_1}\right)^3 
\label{SFQFQConstraint}
\end{align}
where $S_{F_{\rm Q}F_{\rm Q}}^{(1,2)}$ are the contributions to quadratic backaction near $\omega = 0,~\Delta_2$ respectively (cf. Eq.~(\ref{quadraticBA})). These constraints are intuitively clear; large $\bar{n}_2$ reduces the effect of the $\hat{d}_2^{\dagger}\hat{d}_2$ operator in the qubit-cavity coupling relative to the linearised backaction operator, and strong linear cavity detuning ensures the two cavity resonances are well-separated in frequency space. Both these conditions reduce the effect of the quadratic backaction, thereby enabling a stronger dressed intercavity coupling $\widetilde{\lambda}_{12}^{(1)}$. This of course is beneficial for having the nonlinear cavity backaction be the dominant noise source, as described in the discussion following Eq.~(\ref{l12Constraint}). 

The final challenge then is to satisfy the above constraints with experimentally feasible parameter values, allowing close to quantum-limited performance while still being within the weak coupling regime and hence having large gain $\mathcal{G}$. Focusing on realistic parameter choices and requiring weak $\kappa_2$ to suppress the imaginary modification of the Kerr constant (cf. Sec.~\ref{weakCoupling}), we enforce $\kappa_2 \simeq \kappa_1/\sqrt{\mathcal{G}}$. This choice makes both conditions in Eq.~(\ref{SFQFQConstraint}) equivalent; combining the conditions for quantum limited performance in Eqs.~(\ref{l12Constraint}),~(\ref{SFQFQConstraint}), and for the weak coupling regime in Eqs.~(\ref{weakDC}),~(\ref{weakDC2}), we have:
\begin{subequations}\label{grp}
\begin{align}
	\frac{\kappa_1^2}{\sqrt{\mathcal{G}}} < \left(\widetilde{\lambda}_{12}^{(1)}\right)^2 &< \frac{\bar{n}_2(\Delta_2)^2}{\mathcal{G}^{3/2}} \label{finalConstraint} \\ 
	\widetilde{\lambda}_{12}^{(2)} &< 2\Lambda\bar{n}_1~,~ 2\sqrt{ \frac{\Lambda\bar{n}_2}{|\chi_2|}} \label{L12n2Constraint}
\end{align}
\end{subequations}
We reiterate that these conditions are to be satisfied self-consistently. The lower and upper boundaries of inequality~(\ref{finalConstraint}) are where $S_{F_{\rm L}F_{\rm L}}^{\rm NL}$ is comparable to the $S_{F_{\rm L}F_{\rm L}}^{\rm int}$ and $S_{F_{\rm Q}F_{\rm Q}}$ respectively; for ideal performance, we require $\widetilde{\lambda}_{12}^{(1)}$ to be not too close to either boundary, so that $S_{F_{\rm L}F_{\rm L}}^{\rm NL}$ is the dominant noise contribution. We will present an example parameter set that satisfies these constraints in the next section. 

To summarize, near quantum-limited PARI performance requires (i) tuning the relative drive phase $\delta$ to satisfy the reduced backaction condition of Eq.~(\ref{reducedF}), and (ii) choosing a cavity-cavity coupling $\lambda_{12}$ large enough to ensure that the residual nonlinear cavity backaction $S_{F_{\rm L}F_{\rm L}}$ overcomes the intrinsic linear cavity noise, but also not \emph{too large}, so that the quadratic backaction remains small in comparison, while ensuring the nonlinear cavity is not too strongly modified. The balance required forces $\widetilde{\lambda}_{12}^{(1,2)},~\widetilde{\lambda}_{12}^{(2)}$ to satisfy Eqs.~(\ref{finalConstraint}),~(\ref{L12n2Constraint}) respectively.

These constraints allow us to see the advantage of weak nonlinearities in this setup. As mentioned earlier, large $\bar{n}_2$ values are beneficial for quantum-limited performance, as they suppress the quadratic backaction operator. However, $\bar{n}_2$ appears in the dominant modification of the nonlinear cavity bifurcation (cf. Eq.~(\ref{L12n2Constraint})), and cannot be increased indefinitely, \emph{unless} $\lambda_{12}$ is simultaneously decreased so that $\widetilde{\lambda}_{12}^{(2)}$ is held fixed. This, however, raises another potential problem: weakening $\lambda_{12}$ suppresses the gain and the measurement rate, and reduces $S_{F_{\rm L}F_{\rm L}}^{\rm NL}$ relative to $S_{F_{\rm L}F_{\rm L}}^{\rm int}$. The way out is to ensure the dressed coupling $\widetilde{\lambda}_{12}^{(1)}$ is held fixed while $\lambda_{12}$ is decreased; reducing the Kerr nonlinearity $\Lambda$ to increase $\bar{n}_1$ achieves precisely this effect.

\subsection{Optimal performance with realistic parameter values}
\label{optimalPerformance}

We now implement the above guidelines for choosing parameters, focusing on experimentally accessible values that yield near quantum-limited performance with parametric photon number gains  
$\mathcal{G}$ (defined in Eq.~(\ref{parGain})) of $O(100)$, or $O(20~{\rm dB})$;  corresponding results for the gain, measurement rate  and noise are shown in Figs.~\ref{GainVsDelta1} and \ref{MeasVsDelta1}.

As discussed, one ideally needs a large cavity-$1$ photon number $\bar{n}_1$, which requires a relatively weak cavity-$1$ Kerr constant.  We thus take a value $\Lambda = 5 \times 10^{-5} \kappa_1$.  Note that for standard lumped element resonators based on a shunted SQUID, the strength of the Kerr nonlinearity is given by
\begin{align}
	\Lambda/\kappa_1 = \frac{\pi}{2} \frac{R}{R_K},
\end{align}
where $R$ is the effective shunt resistance , and  $R_K = h/e^2$ is the quantum of resistance. For typical setups with $R \sim 50~\Omega$, Kerr nonlinearities on the order of $\Lambda \sim 10^{-2}\kappa_1$ are standard~\cite{Hatridge11}.  Our
chosen value of $\Lambda / \kappa$ thus requires this nonlinearity to be weakened, or `diluted'.  The utility of such weakened nonlinearities, and methods for achieving this, have recently garnered attention \cite{EichlerWallraffEPJ}.
One could for example introduce an additional linear inductance $L_0$ in series with the SQUID (Josephson inductance $L_J$).  This effectively reduces $\Lambda / \kappa$ by a factor of $p^3$, where the participation ratio $p = L_J/(L_0 + L_J)$ can be made small by choosing an appropriately large $L_0$ (see, e.g. Ref.~\cite{EichlerWallraffEPJ}).  Our chosen $\Lambda / \kappa$ thus requires $p \simeq 0.1$.  One must also be careful that the weakened nonlinearity (and consequent larger photon numbers) do not cause higher nonlinearities in the SQUID potential to become relevant.  This condition results in the constraint $Qp \gtrsim 5$, where $Q$ is the nonlinear cavity quality factor~\cite{Manucharyan2007,Schackert2013, Abdo2013, Narla2014}.  Using the parameters in Table~\ref{parChoice} and taking a cavity-$1$ resonance frequency $\omega_{c1} = 6~$GHz yields $Q = 60$ and  $p\simeq 0.1$. One thus satisfies the requirement on $Qp$, while at the same time having a not too-large cavity-$2$ resonance frequency $\omega_{c2} \sim 2\omega_{c1} \sim 12~$GHz.  Finally, we note that weak nonlinearities can also be achieved by replacing the single SQUID with an array of SQUIDs, or by using transmission-line resonators (see Ref.~\cite{Zhou2014, EichlerWallraffEPJ}).


\begin{figure}[t]
\includegraphics[scale = 0.61]{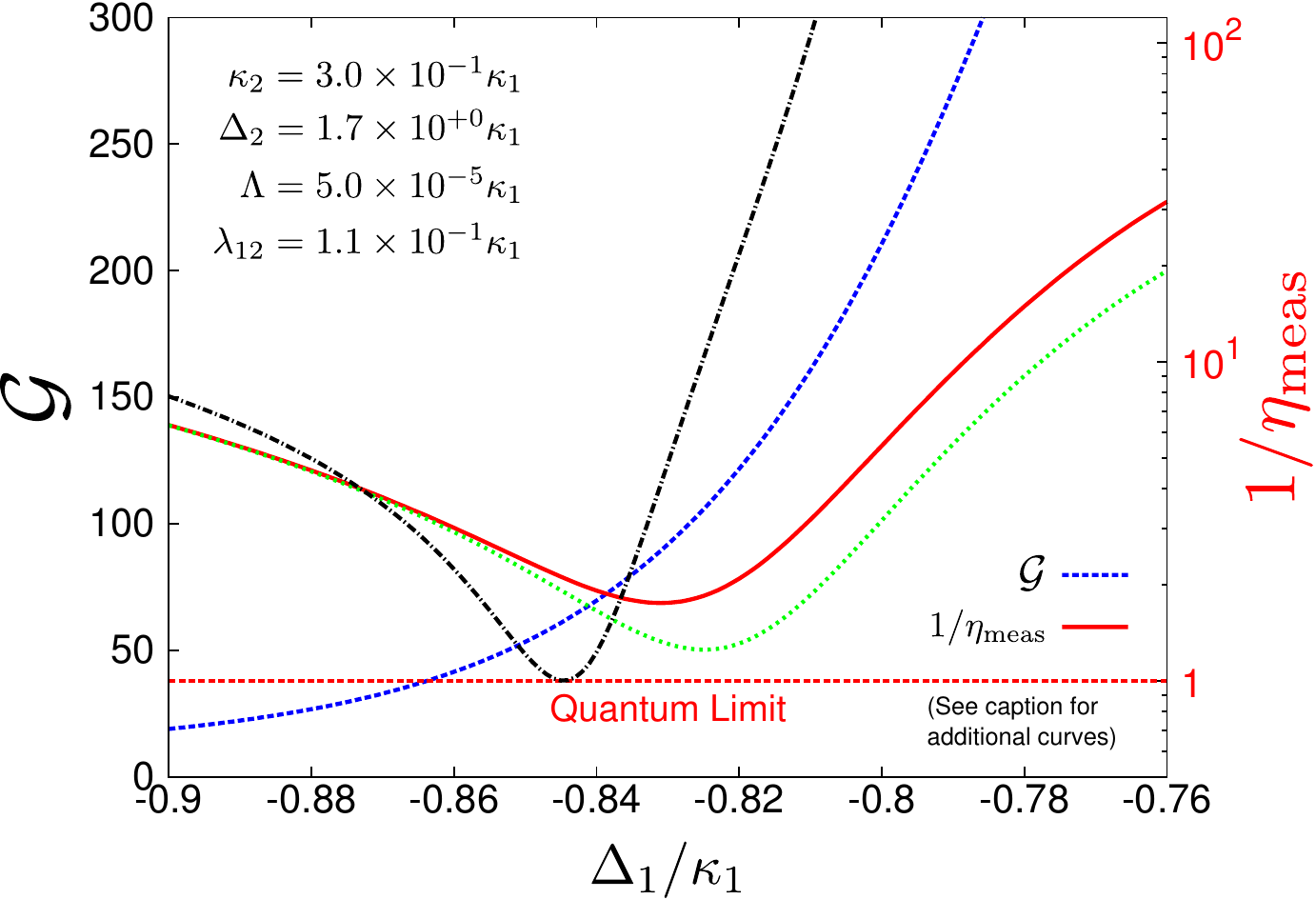}
\caption{Plot of the zero frequency parametric gain $\mathcal{G}$ [cf. Eq.~(\ref{parGain})] (dashed blue, left-hand axis) and the inverse efficiency ratio $1/\eta_{\rm meas}$ (cf. Eq.~(\ref{SzzSff})) (solid red, right hand log axis), as a function of the nonlinear cavity detuning $\Delta_1$. For optimal $\Delta_1 \simeq -0.83\kappa_1$, we obtain $1/\eta_{\rm meas} \simeq 1.75$, with $\mathcal{G} \simeq 100$. The drive parameters are all fixed for this plot: $(\bar{a}_1^{\rm in})^2\!/\kappa_1 = (62.2)^2$, $(\bar{a}_2^{\rm in})^2\!/\kappa_1 = (45.5)^2$, $\delta = 0.40$, $\bar{n}_1 \sim 8000$, $\bar{n}_2 \sim 1200$. The dashed green curve with intermediate minimal backaction-imprecision product corresponds to the limit where $\bar{n}_2 \to \infty$ while $\widetilde{A}$ and $\widetilde{\lambda}_{12}^{(2)}$ are kept fixed; in this limit, the quadratic backaction $S_{F_{\rm Q}F_{\rm Q}}$ (cf. Eq.~(\ref{quadraticBA})) does not contribute. The dot-dashed black curve is the inverse efficiency where we also take $\kappa_2 \to 0$, thus suppressing the intrinsic linear cavity backaction $S_{FF}^{\rm int}$ (cf. Eq.~(\ref{SFLFL})). In this limit, one can exactly reach the quantum limit.}
\label{GainVsDelta1}
\end{figure}

Moving on to the remaining parameters, recall that optimal performance occurs when $\kappa_2 \sim \kappa_1/\sqrt{\mathcal{G}}$; we take $\kappa_2 = 0.3~\kappa_1$. We then implement the optimization conditions. The resulting parameter choices are all shown in Table~\ref{parChoice}.  We take values that are experimentally realistic; more experimentally-challenging choices of parameters would lead to even more optimal performance.  The nonlinear cavity drive strength $\bar{a}_1^{\rm in}$ and detuning $\Delta_1$ are chosen to be close enough to the bifurcation to achieve the required parametric gain. A first order perturbation approach narrows the region in phase space where this bifurcation exists; a numerical optimization routine in $\bar{a}_1^{\rm in}$-$\Delta_1$-$\delta$ space is then implemented to tune these parameters to obtain closest-to-desired detector performance. 

\begin{figure}[t]
\includegraphics[scale = 0.61]{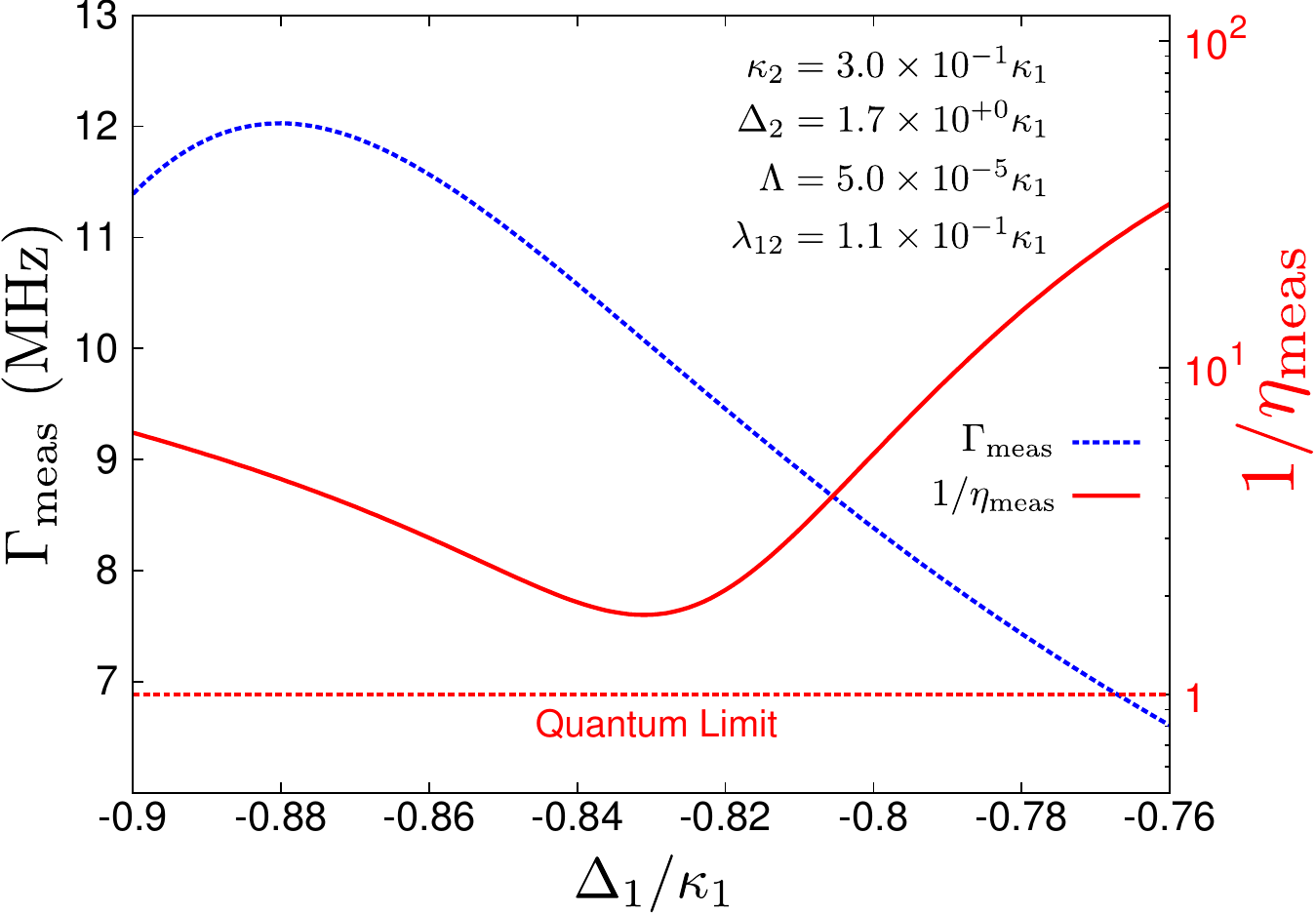}
\caption{Plot of the zero frequency measurement rate $\Gamma_{\mathrm{meas}}$ (dashed blue, left-hand axis) and the inverse efficiency ratio $1/\eta_{\rm meas}$ (solid red, right hand log axis), as a function of the nonlinear cavity detuning $\Delta_1$. For optimal $\Delta_1 = -0.831$, we have $1/\eta_{\rm meas} \simeq 1.75$ and $\Gamma_{\rm meas} \simeq 10~{\rm MHz}$. The drive parameters are the same as for Fig.~\ref{GainVsDelta1}: $(\bar{a}_1^{\rm in})^2\!/\kappa_1 = (62.2)^2$, $(\bar{a}_2^{\rm in})^2\!/\kappa_1 = (45.5)^2$, $\delta = 0.40$, $\bar{n}_1 \sim 8000$, $\bar{n}_2 \sim 1200$. We take $\kappa_1 = 100~$MHz to compute the measurement rate, with the dispersive coupling strength $A = 1~$MHz; the choice for $A$ is justified in the next section. }
\label{MeasVsDelta1}
\end{figure}
In Fig.~\ref{GainVsDelta1}, we plot both the parametric gain $\mathcal{G}$  (dashed blue), as well as the inverse efficiency ratio, $1/\eta_{\rm meas}$ (solid red), both as a function of the nonlinear cavity detuning $\Delta_1$. We find that as predicted, one can both achieve a large effective parametric gain of $\sim 20~$dB (implying a large output signal), and have a backaction noise which is approximately a factor of $1.75$ more than the minimal quantum limited value. The corresponding efficiency is $\eta_{\rm meas} \simeq 0.57$. Had one directly coupled the qubit to the nonlinear cavity, the backaction would exceed the quantum limit by a large factor $\sim 100$.  

In Fig.~\ref{MeasVsDelta1} we plot the measurement rate $\Gamma_{\mathrm{meas}}$ versus $\Delta_1$ for the same parameter set. We focus on the weak dispersive coupling regime, and take $A = 0.01~\kappa_1$. It is possible to obtain measurement rates on the order of $10$ MHz with a nearly quantum-limited detector.

We note that the ability to reach the quantum limit on measurement is restricted by the non-zero intrinsic linear cavity backaction $S_{F_{\rm L}F_{\rm L}}^{\mathrm{int}}$ and the second order backaction $S_{F_{\rm Q}F_{\rm Q}}$. To see this, we include in Fig.~\ref{GainVsDelta1} the dashed green curve showing the inverse efficiency ratio $1/\eta_{\rm meas}$ with $S_{F_{\rm Q}F_{\rm Q}}$ taken to zero ($\bar{n}_2 \to \infty,~\widetilde{A}$ held finite). Also included is the dot-dashed black curve showing this product with $S_{F_{\rm L}F_{\rm L}}^{\mathrm{int}}$ \emph{also} excluded ($\kappa_2 \to 0$), indicating precisely quantum-limited measurement if both noise terms can be suppressed. The relative contributions of each term depends on which boundary of Eq.~(\ref{finalConstraint}) the dressed coupling $\widetilde{\lambda}_{12}^{(1)}$ is closer to.

As mentioned earlier, weak Kerr nonlinearities are advantageous for optimal PARI performance. Using a stronger value of $\Lambda$ limits the cavity intensities $\bar{n}_j$, hence making the quadratic backaction more important. However, the multiple parameters in the PARI provide a way out; adjusting the linear cavity detuning $\Delta_2$ allows us to suppress $S_{F_{\rm Q}F_{\rm Q}}$, albeit at the cost of a reduced measurement rate. In Fig.~\ref{SffProductFinalLambda} in Appendix~\ref{appendixGraphs}, we include an analogous plot to Fig.~\ref{GainVsDelta1}, with a stronger nonlinearity, $\Lambda = 10^{-4}\kappa_1$; we obtain $\eta_{\rm meas} \simeq 0.5$ with $\Gamma_{\rm meas} \sim 1~{\rm MHz}$.

We conclude this section by mentioning the freedom of parameter choices in the PARI scheme, even under the constraints of Eqs.~(\ref{finalConstraint}),~(\ref{L12n2Constraint}). The ability to tune the relative drive phase $\delta$ means that there exist multiple parameter sets allowing optimal PARI performance. This is explicitly shown in Figs.~\ref{productSurf} (a), (b) in Appendix~\ref{appendixGraphs}, where $1/\eta_{\rm meas}$ and $\mathcal{G}$ are analyzed in $\Delta_1$-$\delta$ space; a manifold of optimal operating parameters can clearly be seen to exist.

\subsection{Comparison to conventional circulator-based setup}
\label{compareConvention}

The parametric photon number gain $\mathcal{G}$ (cf. Eq.~(\ref{parGain})) allows a comparison of the amplification performance of the PARI scheme to that of conventional qubit-detector-amplifier setups. It is also useful to develop a meaningful way to compare the noise performance of both schemes in a common framework. 

For the conventional setup, the linear cavity output signal is quantum-limited, but the necessary following amplification stage adds imprecision noise that causes deviations from ideal performance. Fortunately, the qubit is protected from additional backaction due to the use of circulators, and so the backaction noise after amplification, $S_{FF}$, is unchanged from its quantum-limited value. The amplifier boosts the output signal but also makes it noisier; this added noise then modifies the output signal imprecision noise $S_{zz}$ to $\frac{\kappa_1}{2}(1+2\bar{n}_{\rm add})$ (for no noise added by the following amplifier, $\bar{n}_{\rm add} = 0$ and $S_{zz}$ is unchanged). If, following amplification, the backaction-imprecision product deviates from its quantum-limited value by a factor of $x$, then attributing this deviation solely to an increase in the imprecision noise, we can write $S_{zz} = x\frac{\kappa_1}{2} \equiv \frac{\kappa_1}{2}\left(1 + 2\bar{n}_{\mathrm{add}} \right)$, so that $\bar{n}_{\mathrm{add}} = (x-1)/2$. For the parameters used in Figs.~\ref{GainVsDelta1},~\ref{MeasVsDelta1}, we find $x \simeq 1.75$, and the PARI noise performance is therefore equivalent to that of a standard qubit-cavity-amplifier setup where the paramp stage adds $\bar{n}_{\mathrm{add}} \simeq 0.375$ quanta of noise.


\subsection{Validity of linear response theory}

Our analysis assumes operation of the PARI in the limit of weak dispersive coupling, so that the detector can be treated in the linear response regime. It is important to determine whether the parameter choices used in Figs.~\ref{GainVsDelta1},~\ref{MeasVsDelta1} in fact allow the detector response to remain linear in the dispersive coupling strength $A$. A standard approach is to include the exact qubit-cavity dispersive coupling at the level of the classical equations; this coupling leads to a qubit state dependent shift of the linear cavity frequency by $\pm A$, for $\hat{\sigma}_z = \{\uparrow, \downarrow \} = \{1,-1\}$ respectively. We can then compute a classical cavity 1 output homodyne current that knows about $A$:
\begin{equation}
\avg[\hat{I}_{1\sigma_z}](A) = \frac{\kappa_1}{2} \left( e^{i\phi_h} \alpha_{1\sigma_z}(A) + h.c. \right)
\end{equation}
The difference between the currents for the two possible states,
\begin{equation}
\langle \Delta I_1 \rangle = \langle \hat{I}_{1\uparrow} \rangle - \langle \hat{I}_{1\downarrow} \rangle
\label{DeltaI}
\end{equation}
measures how well the detector output discriminates between these states; it is thus a measure of the detector response as a function of $A$. We plot $\langle \Delta I_1 \rangle$ against $A$ in Fig.~\ref{linearRegime} for the same parameters as Figs.~\ref{GainVsDelta1},~\ref{MeasVsDelta1}. For this optimal parameter set, choosing dispersive coupling strengths $A \lesssim 1~$MHz (for $\kappa_1 = 100~$MHz) allows a linear detector response, and our analysis can be safely applied.

\begin{figure}[t]
\includegraphics[scale = 0.62]{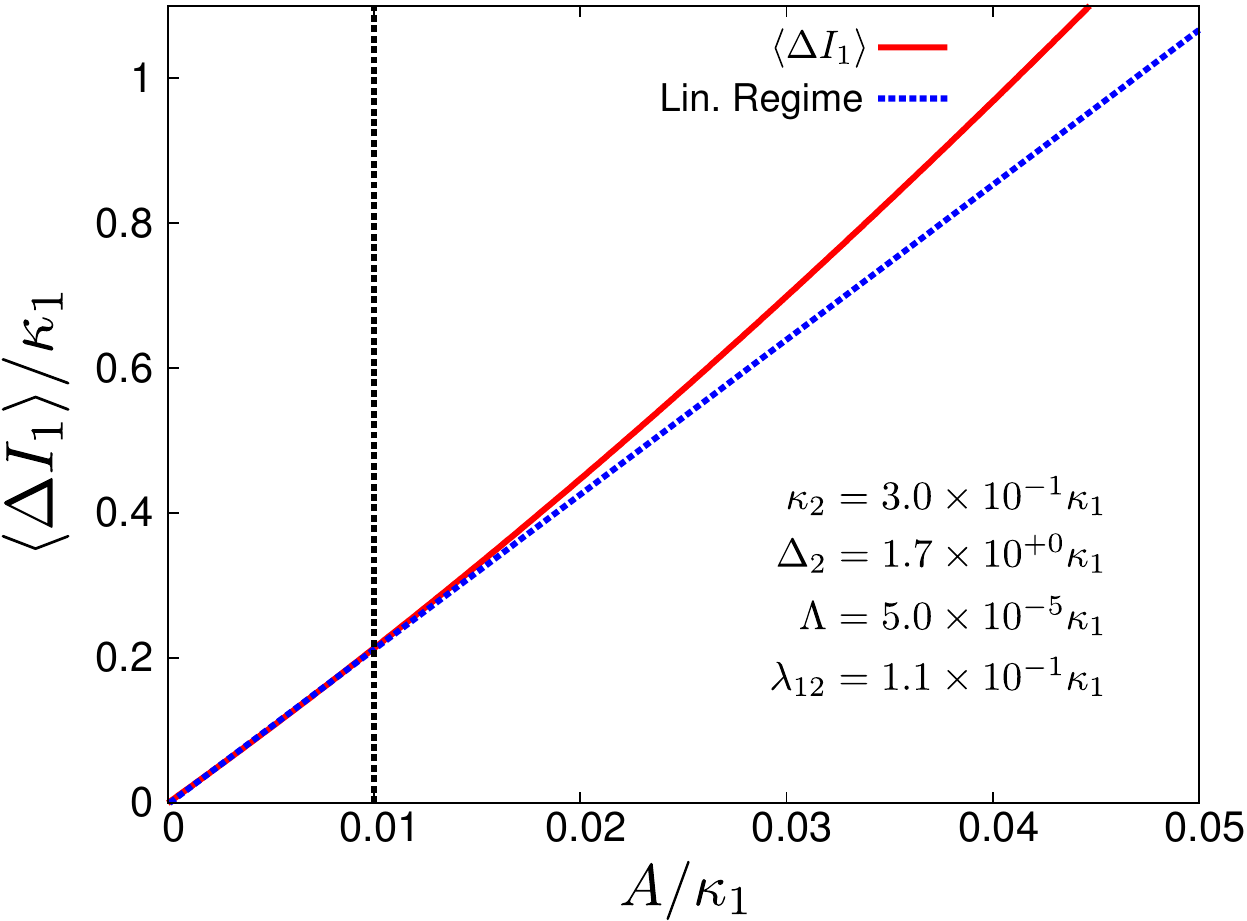}
\caption{Plot of $\langle \Delta I_1 \rangle$ (cf. Eq.~(\ref{DeltaI})) against the dispersive coupling strength $A$. We see that for $A \leq 0.01~\kappa_1$, the system response is approximately linear in $A$ and the results of our analysis hold. We choose a value of $A = 0.01~\kappa_1 = 1~$MHz for Fig.~\ref{MeasVsDelta1} to operate within this linear regime while having a fast measurement rate.}
\label{linearRegime}
\end{figure}

\section{Conclusions}

In this paper, we have detailed the theory of a two-cavity amplifier for qubit state measurement in the linear response regime. The setup, where the linear cavity housing the qubit is coupled directly to the nonlinear cavity used for measurement, can be mapped to a single modified nonlinear cavity coupled to the qubit via a tunable coupling. Adhering to a weak intercavity coupling regime ensures retention of the nonlinear cavity's bifurcation, and hence a large amplifier gain. We find further that the linear cavity's response modifies the amplifier's properties. Crucially, the modified qubit-cavity coupling can be tuned to shield the qubit from linear measurement backaction, while the multiple parameter choices allow suppressing the quadratic backaction, all while preserving the large gain. Therefore, quantum-limited qubit state measurement that escapes a single nonlinear cavity detector within weak coupling is possible with the PARI device.

\section{Acknowledgements}

We thank Andrew Eddins, David Toyli, and Eli-Levenson Falk for many useful discussions. This work was supported by NSERC and the Army Research Office under Grant \#: W911NF-14-1-0078.

\begin{center}
\noindent\rule{4cm}{0.4pt}
\end{center}


\appendix

\section{Derivation of intercavity coupling Hamiltonian}
\label{couplingDerivation}

For the setup shown in Fig.~\ref{boxSchematic}~(b), the Josephson energy of the nonlinear circuit's SQUID can be written in terms of the flux
threading the SQUID loop,
\begin{equation}
\hat{\Phi}_{\mathrm{ext}} = \Phi_B + M\hat{I}_2
\end{equation}
where $\Phi_B$ is a static flux bias. The second term contains the contribution to the flux due to the coupling with the linear cavity, and is proportional to the current $I_2$ flowing in the linear resonator. $M$ is the mutual inductance of the coupled cavity system. The physical Hamiltonian of the nonlinear cavity in Fig.~\ref{boxSchematic}~(b) is therefore:
\begin{equation}
\hat{H}_1^{\mathrm{sys}} = \frac{\hat{Q}_1^2}{2C_1}  - E_J[\hat{\Phi}_{\mathrm{ext}}]~\cos \left( 2\pi
\frac{\hat{\Phi}_1}{\Phi_0} \right)
\end{equation}
where $\hat{\Phi},\hat{Q}$ are the canonical coordinates describing the nonlinear cavity with shunt capacitance $C_1$, and $\Phi_0 = h/2e$ is the magnetic flux quantum. The first term includes the capacitive energy of the circuit, while the second
is the Josephson energy term providing the nonlinearity and the intercavity coupling. This expression can be simplified via Taylor expansion, keeping terms up to and including $\hat{\Phi}_1^4$, ignoring constant terms and the term proportional only to $\hat{I}_2$ (these are accounted for in the total drive on the linear cavity). As a result, we obtain the quadratic- and quartic-in-$\hat{\Phi}_1$ terms that form the nonlinear cavity Hamiltonian with a Kerr type nonlinearity, as is standard, together with a term depending on both $\Phi_1$ and $\hat{I}_2$; we focus on this intercavity coupling term:
\begin{equation}
\hat{H}_{12} = \Bigg\{ \frac{M}{2}\left( \frac{2\pi}{\Phi_0} \right)^2\!\!\frac{d E_J}{d \Phi_{\mathrm{ext}}} \Bigg|_{\Phi_B}\!  \Bigg\}\hat{I}_2 \hat{\Phi}_1^2
\label{h12orig}
\end{equation}
To generalize and simplify the analysis, we now introduce the usual creation and annihilation operators for both cavities,
\begin{align}
\hat{a}_j =& \frac{1}{\sqrt{2\hbar L_j \omega_{cj}}} \hat{\Phi}_j + i\frac{1}{\sqrt{2\hbar C_j \omega_{cj}}}
\hat{Q}_j  \nonumber \\
\hat{a}_j^{\dagger} =& \frac{1}{\sqrt{2\hbar L_j \omega_{cj}}} \hat{\Phi}_j - i\frac{1}{\sqrt{2\hbar C_j \omega_{cj}}}\hat{Q}_j 
\label{ops}
\end{align}
where $\omega_{cj} = 1/\sqrt{L_jC_j}$ is the natural frequency of cavity~$j$ ($j = 1,2$). Using the fact that $\hat{I}_2 = \hat{\Phi}_2/L_2$, we can rewrite the intercavity coupling term in Eq.~(\ref{h12orig}) in terms of creation and annihilation operators to get the form written in Eq.~(\ref{H12}). From there, the intercavity coupling strength $\lambda_{12}$ can be extracted in terms of physical quantities,
\begin{equation}
\lambda_{12} = 2\omega_{c1}\sqrt{\frac{  \hbar \omega_{c2}}{2 L_2} }~\left[ \frac{M}{E_J[\Phi_B]} \frac{d E_J}{d \Phi_{\mathrm{ext}}} \Bigg|_{\Phi_B} \right]
\label{lambda12}
\end{equation}
To obtain a numerical estimate, we use the explicit form for the Josephson energy $E_J[\Phi] = E_0 \left| \cos \left( \frac{\pi \Phi}{\Phi_0} \right) \right|$, and rewrite $L_2$ in terms of the characteristic impedance $Z_2 = \sqrt{L_2/C_2}$ and $\omega_{c2}$, hence simplifying the expression for $\lambda_{12}$,
\begin{equation}
\lambda_{12} =  \left[M \cdot (\omega_{c1}\omega_{c2})\cdot \frac{1}{R_K}\right]  \sqrt{\frac{R_K }{Z_2}} \sqrt{
4\pi}~\tan \left( \frac{\pi \Phi_B}{\Phi_0} \right)
\end{equation}i
where $R_K = h/e^2 = 2\Phi_0/e$ is the quantum of resistance. Taking $\omega_{c1} = (2\pi)6~$GHz, $\omega_{c2} = (2\pi)12~$GHz, choosing $Z_2 \simeq Z_0$, where $Z_0$ is the impedance of free space, and noting that the angular factor will be $O(1)$ for the static bias fluxes used in such setups, we obtain,
\begin{equation}
\lambda_{12} \sim  \left( \frac{M}{\mathrm{pH}} \right)~10^{-1}~\mathrm{MHz}
\end{equation}
From the considered parameter choices (cf. Table~\ref{parChoice}), we have a value of $\lambda_{12} \simeq 1.1~$MHz, which would correspond to a mutual inductance $M \simeq 11~\mathrm{pH}$.


\section{Classical equations of motion}
\label{appendixClassical}

We consider the full Heisenberg equations for the two cavities, Eq.~(\ref{heisenbergA}), and compute the classical equations of motion
assuming a steady state coherent solution of the form indicated in Eq.~(\ref{modes}), without the quantum operators. Imposing the driving frequency condition [cf. Eq.~(\ref{driveCond})] followed by a rotating wave approximation allows one to easily obtain the following classical equations of motion for the nonlinear and linear cavities respectively:
\begin{align}
\alpha_1 \left[-i(\Delta_1 + 2\Lambda |\alpha_1|^2) + \kappa_1/2 \right] =& - \sqrt{\kappa_1}~\bar{a}_{1}^{\mathrm{in}} -i\lambda_{12}
\alpha_2 \alpha_1^* \nonumber  \\
\alpha_2 \left[-i\Delta_2 + \kappa_2/2 \right] =& - \sqrt{\kappa_2}~\bar{a}_{2}^{\mathrm{in}}e^{i \delta} -i\frac{\lambda_{12}}{2} \alpha_1^2
\label{alphaApp}
\end{align}
The second equation can be used to eliminate $\alpha_2$ to obtain a complex nonlinear equation for $\alpha_1$,
\begin{align}
\alpha_1 \left[-i(\Delta_1 + 2\Lambda |\alpha_1|^2) + \kappa_1/2 \right] =& - \sqrt{\kappa_1}~\bar{a}_{1}^{\mathrm{in}} \nonumber \\
+i\lambda_{12} (\sqrt{\kappa_2}~\bar{a}_{2}^{\mathrm{in}}e^{i\delta}~\chi_{2}) \alpha_1^* &-\frac{\lambda_{12}^2}{2} ( \chi_{2}) \alpha_1^2 \alpha_1^* 
\label{classEOMA1}
\end{align}
The first line in the equation above is just the classical equation of motion for an individual nonlinear
cavity, while the second line represents the modifications due to the coupling to the linear cavity. We can
rewrite the modifications proportional to $\lambda_{12}^2$ as changing the Kerr constant,
\begin{align}
\alpha_1 \left[-i(\Delta_1 + 2\widetilde{\Lambda} |\alpha_1|^2) + \kappa_1/2 \right] -i\eta\alpha_1^* =& -
\sqrt{\kappa_1}~\bar{a}_{1}^{\mathrm{in}} 
\end{align}
where the modified Kerr constant $\widetilde{\Lambda}$ is as defined in Eq.~(\ref{complexKerr}), and the emergent
parametric driving strength $\eta$ is as defined in Eq.~(\ref{eta}). Eq.~(\ref{classEOMA1}) mixes $\alpha_1$ and
$\alpha_1^*$, so that the phase of $\alpha_1$ plays an important role. However, we now note that it is possible to write down the above
equation and its complex conjugate as a matrix system:
\begin{equation}
\begin{bmatrix}
\kappa_1/2 -i\zeta & -i\eta \\
i\eta^* & \kappa_1/2 + i\zeta^*  \\
\end{bmatrix}
\begin{pmatrix}
\alpha_1 \\
\alpha_1^*
\end{pmatrix} = -\sqrt{\kappa_1} 
\begin{pmatrix}
\bar{a}_1^{\mathrm{in}} \\
\bar{a}_1^{\mathrm{in}}
\end{pmatrix}
\end{equation}
where $\zeta$ is given by
\begin{equation}
\zeta = \Delta_1 + 2\widetilde{\Lambda} |\alpha_1|^2
\end{equation}
Inverting the 2-by-2 matrix above allows an expression for $\alpha_1$ that does not include explicit reference to its phase; squaring this expression allows one to obtain Eq.~(\ref{quinticClassical}). The equation can be solved numerically for the nonlinear cavity photon number profiles indicated in Fig.~\ref{n1Profiles}.

The purely real equation is also more tractable for a perturbative expansion in the weak intercavity coupling strength. We expand the detuning, drive strength, and nonlinear cavity occupation around the unmodified nonlinear cavity bifurcation $(\Delta_{1,0}, \bar{a}_{1,0}^{\mathrm{in}})$ (where $d\bar{n}_1/d\Delta_1 \to \infty$)~\cite{Laflamme2011},
\begin{align}
\Delta_1 &= \Delta_{1,0} + \frac{\lambda_{12}}{\kappa_1}\cdot \Delta_{1,1} \nonumber \\
\bar{a}_1^{\mathrm{in}} &= \bar{a}_{1,0}^{\mathrm{in}} + \frac{\lambda_{12}}{\kappa_1}\cdot \bar{a}_{1,1}^{\mathrm{in}}
\nonumber \\
\bar{n}_1 &= \bar{n}_{1,0} + \frac{\lambda_{12}}{\kappa_1}\cdot \bar{n}_{1,1} 
\label{pertExpansion}
\end{align}

To retain the bifurcation, we impose that $\Delta_{1,1},~\bar{a}_{1,1}^{\mathrm{in}},~\bar{n}_{1,1}$ be chosen such that the $O(\lambda_{12})$ modifications to
both the classical nonlinear equation and the diverging derivative at bifurcation $d\bar{n}_1/d\Delta_1$, vanish. This yields a modified set of detuning and driving strengths describing the new position of the bifurcation:
\begin{subequations}\label{grp}
\begin{align}
\Delta_1 &= \Delta_{1,0} + \lambda_{12} \sqrt{\kappa_2}~\bar{a}_2^{\mathrm{in}}|\chi_2|\left[ \cos (\delta + \phi_2) + \sqrt{3} \sin (\delta + \phi_2 ) \right]   \nonumber \\
\bar{a}_1^{\mathrm{in}} &= \bar{a}_{1,0}^{\mathrm{in}} + \lambda_{12}\sqrt{\kappa_2}~\bar{a}_2^{\mathrm{in}}~|\chi_2| \left[\sin (\delta +
\phi_2)\left( \sqrt{3}/2\Lambda \right)^{1/2} \right] 
\label{a11}
\end{align} 
\end{subequations}
where $\phi_2$ is defined via $\chi_2 = |\chi_2|\exp({i\phi_2})$ as before. An important observation - borne out by the numerics - is that the critical detuning and driving strength values are affected by the relative drive phase $\delta$ in particular, indicating that tuning $\delta$ nontrivially modifies the bifurcation physics.


\section{Effective nonlinear cavity}
\label{appendixEffectiveM}

The set of four coupled Heisenberg equations of motion in the displaced frame, Eqs.~(\ref{heisenbergD}), form a four-dimensional system. It is clearest to solve this system by expressing it in the canonical quadrature basis; this transform is carried out using the matrix $\mathbf{T}_c$,
\begin{equation}
\begin{pmatrix}
\hat{x}_1 \\
\hat{p}_1 \\
\hat{x}_2 \\
\hat{p}_2 \\
\end{pmatrix} = 
\mathbf{T}_c
\begin{pmatrix}
\hat{d}_1 \\
\hat{d}_1^{\dagger} \\
\hat{d}_2 \\
\hat{d}_2^{\dagger} \\
\end{pmatrix} = 
\frac{1}{ \sqrt{2}} \begin{bmatrix}
1 & 1 & 0 & 0 \\
-i & i & 0 & 0 \\
0 & 0 & 1 & 1 \\
0 & 0 & -i & i \\
\end{bmatrix}
\begin{pmatrix}
\hat{d}_1 \\
\hat{d}_1^{\dagger} \\
\hat{d}_2 \\
\hat{d}_2^{\dagger} \\
\end{pmatrix} 
\label{dToCan}
\end{equation} \\
Eq.~(\ref{dToCan}) is also used to transform the input noise operators $\hat{\xi}_j$ to the canonical
basis, with the replacements $(\hat{d}_j,\hat{d}_j^{\dagger}) \to (\hat{\xi}_j,\hat{\xi}_j^{\dagger}),~(\hat{x}_j,\hat{p}_j)
\to (\hat{x}_j^{\mathrm{in}},\hat{p}_j^{\mathrm{in}})$. Following this, the full four-dimensional system in the canonical basis
can be written in Fourier space as,
\begin{widetext}
\begin{equation}
(i \omega \mathbf{1} + \mathbf{M}_c) \begin{pmatrix}
\hat{x}_1[\omega] \\
\hat{p}_1[\omega]  \\
\hat{x}_2[\omega] \\
\hat{p}_2[\omega]  \\
\end{pmatrix} = \begin{pmatrix}
\sqrt{\kappa_1}~\hat{x}_{1}^{\mathrm{in}}[\omega] \\
\sqrt{\kappa_1}~\hat{p}_{1}^{\mathrm{in}}[\omega]  \\
\sqrt{\kappa_2}~\hat{x}_{2}^{\mathrm{in}}[\omega] \\
\sqrt{\kappa_2}~\hat{p}_{2}^{\mathrm{in}}[\omega] \\
\end{pmatrix}
+
\sqrt{2}
\begin{pmatrix}
0 \\
0 \\
-\cos \left( \frac{\phi_{\varg}}{2} + \frac{3\pi}{4} \right) \\
\sin \left( \frac{\phi_{\varg}}{2} + \frac{3\pi}{4} \right) 
\end{pmatrix}\!\hat{B}[\omega]
\label{cavEOMFS1}
\end{equation}
\end{widetext}
Here $\hat{B}[\omega] = \widetilde{A}\hat{\sigma}_z[\omega]$ describes the qubit's frequency-dependent driving force on the linear cavity via the linearised dispersive coupling, and the susceptibility matrix $\mathbf{M}_c$ is:
\begin{equation}
\mathbf{M}_{c} = 
\begin{bmatrix}
\mathbf{M}_1 & -\lambda_{12}\sqrt{\bar{n}_1}~\mathbf{R}(-\mu_{12}-\pi/2) \\
\lambda_{12}\sqrt{\bar{n}_1}~\mathbf{R}(\mu_{12} + \pi/2) & \mathbf{M}_2   \\
\end{bmatrix}
\end{equation}
where the nonlinear cavity response matrix $\mathbf{M}_1$ and linear cavity response matrix $\mathbf{M}_2$ take the forms:
\begin{equation}
\mathbf{M}_1 = 
\begin{bmatrix}
\varg_{\mathrm{eff}} - \frac{\kappa_1}{2} & -\widetilde{\Delta}_1 \\
\widetilde{\Delta}_1 & -\varg_{\mathrm{eff}} -  \frac{\kappa_1}{2} 
\end{bmatrix} ,~
\mathbf{M}_2 = 
\begin{bmatrix}
-\frac{\kappa_2}{2} & -\Delta_2 \\
\Delta_2 & -\frac{\kappa_2}{2} 
\end{bmatrix}
\label{M1M2}
\end{equation}
The off-diagonal matrices - the cross-terms - describe the intercavity coupling; in particular, these appear in the form
of an effective rotation defined by the matrix $\mathbf{R}(\theta)$:
\begin{equation}
\mathbf{R}(\theta) = 
\begin{bmatrix}
\cos \theta & \sin \theta \\
-\sin \theta & \cos \theta
\end{bmatrix};~~~~~~\mathbf{R}^{-1}(\theta) = \mathbf{R}(-\theta)
\label{rot}
\end{equation}
The rotation is characterised by the angle $\mu_{12}$ defined in Eq.~(\ref{mu}). It is convenient to use the block matrix structure to expand Eq.~(\ref{cavEOMFS1}) into equations for both cavity modes separately:
\begin{widetext}
\begin{subequations}\label{grp}
\begin{align}
(i\omega\mathbf{1} + \mathbf{M}_1)
\begin{pmatrix}
\hat{x}_1[\omega] \\
\hat{p}_1[\omega]
\end{pmatrix}
- \lambda_{12}\sqrt{\bar{n}_1}~\mathbf{R}\left(-\mu_{12}-\pi/2\right)
\begin{pmatrix}
\hat{x}_2[\omega] \\
\hat{p}_2[\omega]
\end{pmatrix} =
\sqrt{\kappa_1}
\begin{pmatrix}
\hat{x}_1^{\mathrm{in}}[\omega] \\
\hat{p}_1^{\mathrm{in}}[\omega]
\end{pmatrix} \label{x1p1} \\
(i\omega\mathbf{1} + \mathbf{M}_2)
\begin{pmatrix}
\hat{x}_2[\omega] \\
\hat{p}_2[\omega]
\end{pmatrix}
+ \lambda_{12}\sqrt{\bar{n}_1}~\mathbf{R}\left(\mu_{12}+\pi/2\right)
\begin{pmatrix}
\hat{x}_1[\omega] \\
\hat{p}_1[\omega]
\end{pmatrix} =
\sqrt{\kappa_2}
\begin{pmatrix}
\hat{x}_2^{\mathrm{in}}[\omega] \\
\hat{p}_2^{\mathrm{in}}[\omega]
\end{pmatrix} \label{x2p2}
\end{align}
\end{subequations}
\end{widetext}
Note that we ignore the coupling to the qubit for the moment for clarity. A crucial part of our analysis involves solving Eq.~(\ref{x2p2}) for the linear cavity quadratures and substituting the results into Eq.~(\ref{x1p1}), to obtain a set of equations for the cavity-1 quadratures only. We thus obtain Eq.~(\ref{2DSys}), an effective two-dimensional system with the linear cavity eliminated:
\begin{align}
\begin{pmatrix}
\hat{x}_1[\omega] \\
\hat{p}_1[\omega]
\end{pmatrix}
&=\mathbf{M}_{\mathrm{eff}}^{-1}[\omega] 
\begin{pmatrix}
\hat{x}_e^{\mathrm{in}}[\omega] \\
\hat{p}_e^{\mathrm{in}}[\omega]
\end{pmatrix}
\label{2DSysA}
\end{align}
The effective response matrix $\mathbf{M}_{\mathrm{eff}}[\omega]$ can be written in terms of the original nonlinear cavity response matrix $\mathbf{M}_1$ together with the self-energy matrix introduced in Eq.~(\ref{MeffW}),
\begin{align}
\mathbf{M}_{\mathrm{eff}}[\omega] &\equiv \mathbf{M}_1[\omega] - i\left(\widetilde{\lambda}_{12}^{(1)}\right)^2 \!\! \mathbf{\Sigma}[\omega]  \nonumber \\
\mathbf{\Sigma}[\omega] &=
\frac{i}{2}
\begin{bmatrix}
 (\chi_{2}[\omega] + \chi_{2}^*[-\omega]) & i(\chi_{2}[\omega] -\chi_{2}^*[-\omega]) \\
-i (\chi_{2}[\omega] - \chi_{2}^*[-\omega]) &  (\chi_{2}[\omega] + \chi_{2}^*[-\omega])
\end{bmatrix}
\label{selfEnergy}
\end{align}
where the full linear cavity susceptibility takes the form
\begin{equation}
\chi_{2}[\omega] = \left( -i\omega + \kappa_2/2 \pm i\Delta_2 \right)^{-1}
\label{chi2Full}
\end{equation}
and $\mathbf{M}_2[\omega] = i\omega\mathbf{1} + \mathbf{M}_2$. $\mathbf{M}_{\mathrm{eff}}[\omega]$ is given explicitly by,
\begin{equation}
\mathbf{M}_{\mathrm{eff}}[\omega] = 
\begin{bmatrix}
\varg_{\mathrm{eff}} - \frac{\kappa_{\mathrm{eff}}}{2}[\omega] & -\widetilde{\Delta}_{\mathrm{eff}}[\omega] \\
\widetilde{\Delta}_{\mathrm{eff}}[\omega] & -\varg_{\mathrm{eff}} - \frac{\kappa_{\mathrm{eff}}}{2}[\omega] 
\end{bmatrix}
\label{MeffFull}
\end{equation}
where the effective nonlinear cavity detuning, damping, and parametric interaction strength are:
\begin{align}
\widetilde{\Delta}_{\mathrm{eff}}[\omega] &= \widetilde{\Delta}_1 + \left(\widetilde{\lambda}_{12}^{(1)}\right)^2 \! \frac{i}{2}\left( \chi_{2}[\omega] - \chi_{2}^*[-\omega] \right)  \nonumber \\
\kappa_{\mathrm{eff}}[\omega] &= \kappa_1+ \left(\widetilde{\lambda}_{12}^{(1)}\right)^2 \!\! \left( \chi_{2}[\omega]+ \chi_{2}^*[-\omega] \right) \nonumber \\
\varg_{\mathrm{eff}} &= \left|\varg_1 - \widetilde{\lambda}_{12}^{(2)}e^{i\mu_{12}}\right|
\label{paramseFull}
\end{align}
Finally, the drive vector on the nonlinear cavity is also modified, as describe in Eqs.~(\ref{2DSys}); this can then be expressed as:
\begin{widetext}
\begin{equation}
\begin{pmatrix}
\hat{x}_e^{\mathrm{in}}[\omega] \\
\hat{p}_e^{\mathrm{in}} [\omega]
\end{pmatrix}
= \sqrt{\kappa_1} 
\begin{pmatrix}
\hat{x}_{1}^{\mathrm{in}}[\omega]\\
\hat{p}_{1}^{\mathrm{in}}[\omega]
\end{pmatrix}
-\widetilde{\lambda}_{12}^{(1)}\left\{\mathbf{R}\left(\mu_{12} + \pi/2 \right]\mathbf{M}_2[\omega] \right\}^{-1}\!\!\sqrt{\kappa_2}
\begin{pmatrix}
\hat{x}_{2}^{\mathrm{in}}[\omega] \\
\hat{p}_{2}^{\mathrm{in}}[\omega]
\end{pmatrix}
\label{driveEff}
\end{equation}
\end{widetext}
We now reintroduce the coupling to the qubit in Eqs.~(\ref{cavEOMFS1}); the only change is to the modified drive on the effective system. It can be obtained from Eq.~(\ref{driveEff}) using the replacement:
\begin{equation}
\sqrt{\kappa_2}
\begin{pmatrix}
\hat{x}_{2}^{\mathrm{in}} \\
\hat{p}_{2}^{\mathrm{in}}
\end{pmatrix} \to 
\sqrt{\kappa_2}
\begin{pmatrix}
\hat{x}_{2}^{\mathrm{in}} \\
\hat{p}_{2}^{\mathrm{in}}
\end{pmatrix} + 
\sqrt{2}
\begin{pmatrix}
-\cos \left( \frac{\phi_g}{2} + \frac{3\pi}{4} \right)  \\
\sin \left( \frac{\phi_g}{2} + \frac{3\pi}{4} \right) 
\end{pmatrix}
\!\hat{B}[\omega]
\label{driveQubit}
\end{equation}
(where the frequency labels of the quadratures are suppressed).


The matrix $\mathbf{M}_{\mathrm{eff}}[\omega]$ (cf. Eq.~(\ref{MeffFull})) bears similarity to the susceptibility matrix for a driven, damped nonlinear cavity with no coupling to a linear cavity. The full system is solved easily by first diagonalizing $\mathbf{M}_{\mathrm{eff}}[\omega]$,
\begin{equation}
\mathbf{M}_{\mathrm{eff}}[\omega] = -\mathbf{V}_{e} 
\begin{bmatrix}
(\chi_{e-}[\omega])^{-1} & 0 \\
0 & (\chi_{e+}[\omega])^{-1}  \\
\end{bmatrix} \mathbf{V}_{e}^{-1}
\label{m1}
\end{equation}
where $\mathbf{V}_{e}$, the matrix of eigenvectors of $\mathbf{M}_{\mathrm{eff}}[\omega]$, is:
\begin{equation}
\mathbf{V}_{e} = 
\begin{bmatrix}
\cos(\theta_{e}/2) & \sin(\theta_{e}/2) \\
\sin(\theta_{e}/2) & \cos(\theta_{e}/2) \\
\end{bmatrix}
\label{v1e}
\end{equation}
$\theta_e$ is defined in Eq.~(\ref{thetae}), and the frequency-dependent effective nonlinear cavity susceptibilities are simply $\chi_{e\pm}[\omega]^{-1} = \left( -i\omega + \chi_{e\pm}^{-1} \right)$, with $\chi_{e\pm}$ defined in Eq.~(\ref{chiEP}),~(\ref{chiEM}). 

The analysis is most transparent in the rotated basis of amplified and squeezed quadratures, $\hat{X}_e, \hat{P}_e$ respectively, as discussed earlier. Using the transformation:
\begin{align} 
\begin{pmatrix}
\hat{X}_{e} \\
\hat{P}_{e} \\
\end{pmatrix} = 
\mathbf{R}(\theta_e/2)
\begin{pmatrix}
\hat{x}_{1} \\
\hat{p}_{1} 
\end{pmatrix}  ~;~ 
\begin{pmatrix}
\hat{X}_{e}^{\mathrm{in}} \\
\hat{P}_{e}^{\mathrm{in}} 
\end{pmatrix}  = 
\mathbf{R}(\theta_e/2)
\begin{pmatrix}
\hat{x}_{e}^{\mathrm{in}} \\
\hat{p}_{e}^{\mathrm{in}} 
\end{pmatrix} 
\label{Te}
\end{align}
where all quadratures are understood to be frequency dependent, the effective two-dimensional system of Eq.~(\ref{2DSys}) can be written in this rotated basis as:
\begin{equation}
\begin{pmatrix}
\hat{X}_{e}[\omega] \\
\hat{P}_{e}[\omega] 
\end{pmatrix} = 
\mathbf{M}^{-1}_{\mathrm{eff}}[\omega] 
\begin{pmatrix}
\hat{X}_{e}^{\mathrm{in}}[\omega] \\
\hat{P}_{e}^{\mathrm{in}}[\omega]
\end{pmatrix}
\label{effSys}
\end{equation}
where $\mathbf{M}_{\mathrm{eff}}^{-1}[\omega]$ is given in terms of $\chi_{e\pm}[\omega]$ by:
\begin{equation}
\mathbf{M}_{\mathrm{eff}}^{-1}[\omega] = 
\begin{bmatrix}
-\chi_{e-}[\omega] & \left(\chi_{e-}[\omega] - \chi_{e+}[\omega] \right) \tan \theta_e \\
0 & -\chi_{e+}[\omega] 
\end{bmatrix}
\label{MeffSusc}
\end{equation}
The effective cavity output quadratures are then simply obtained by multiplying out Eqs.~(\ref{effSys}), leading to the frequency-dependent versions of Eqs.~(\ref{Xe}),~(\ref{Pe}) (acquired by adding frequency labels to all quantities).


\section{Backaction force noise calculations}
\label{appendixF}

\subsubsection{Linearised backaction calculation}
We begin by elaborating on the calculation of the linearised backaction, and start again by express the linearised backaction force operator $\hat{F}_{\rm L}$ (cf. Eq.~(\ref{Fop})) in the rotated quadrature basis described in Appendix~\ref{appendixEffectiveM}. This is achieved by eliminating the linear cavity modes from $\hat{F}_{\rm L}$, thereby expressing it in terms of the nonlinear cavity modes only. Upon transforming to the $\hat{X}_e,~\hat{P}_e$ basis, $\hat{F_{\rm L}}$ takes the form given in Eq.~(\ref{FE}). 
The $\hat{F}_{\rm L}^{\rm int}[\omega]$ operator is given in terms of the linear cavity input noise operators $\hat{\xi}_2[\omega]$ \emph{only}, and is in particular independent of $\lambda_{12}$:
\begin{align}
\hat{F}_{\rm L}^{\rm int}[\omega] = \sqrt{\kappa_2}~\widetilde{A}~\Big(&\chi_2[\omega]e^{i(\phi_g/2 - 3\pi/4)} \hat{\xi}_2[\omega] \nonumber \\
+ &\chi_2^*[-\omega]e^{-i(\phi_g/2 - 3\pi/4)} \hat{\xi}_2^{\dagger}[\omega]  \Big)
\label{fINT}
\end{align}
The $\lambda_{12}$ dependent backaction terms involve $f_X[\omega]$, defined in Eq.~(\ref{FXFull}), and $f_P[\omega]$, which is defined as:
\begin{equation}
f_P[\omega] = i\frac{\widetilde{A}}{\sqrt{2}}~\left[\chi_2^*[-\omega]e^{i(\mu_{12} - \nu[\omega])} - \chi_2[\omega]e^{-i(\mu_{12}-\nu[\omega])} \right]
\label{FP} 
\end{equation}
We now move on to characterising the noise spectral density of the linearised backaction force. For calculations at non-zero frequencies, we compute the symmetrized noise spectral density, $\bar{S}_{FF}[\omega]$, where:
\begin{align}
\bar{S}_{FF}[\omega] = \frac{1}{2}\left(S_{FF}[\omega] + S_{FF}[-\omega] \right)
\end{align}
The unsymmetrized spectral density $S_{FF}[\omega]$, given by Eq.~(\ref{SFFdef}), can be computed using an application of the Wiener-Khinchin theorem, and the form of $\hat{F_{\rm L}}[\omega]$. It is simple to relate $\hat{F}_{\rm L}$ to the vacuum drives on both cavities, $\hat{X}_j^{\mathrm{in}},\hat{P}_j^{\mathrm{in}}$ in the rotated basis using the solutions to Eqs.~(\ref{effSys}). $S_{FF}[\omega]$, and hence $\bar{S}_{FF}[\omega]$, can then be calculated using the usual correlation functions for these vacuum noise drives at zero temperature,
\begin{align}
\langle \hat{X}_{j}^{\mathrm{in}}[\omega]\hat{X}_{j}^{\mathrm{in}}[\omega'] \rangle &= \pi
\delta(\omega+\omega')
\nonumber \\
\langle \hat{P}_{j}^{\mathrm{in}}[\omega]\hat{P}_{j}^{\mathrm{in}}[\omega'] \rangle &= \pi
\delta(\omega+\omega')
\nonumber \\
\langle \hat{X}_{j}^{\mathrm{in}}[\omega]\hat{P}_{j}^{\mathrm{in}}[\omega'] \rangle &= i\pi
\delta(\omega+\omega')
\label{corrs}
\end{align}
\subsubsection{Quadratic backaction calculation}

The second backaction source is the quadratic backaction operator $\hat{F}_{\rm Q}$ (cf. Eq.~(\ref{Fop})). The effect of this quadratic backaction can be computed from the relevant autocorrelation function $G_{F_{\rm Q}F_{\rm Q}}(t)$,
\begin{align}
&G_{F_{\rm Q}F_{\rm Q}}(t) =  \langle \langle \hat{F}_{\rm Q}(t)\hat{F}_{\rm Q}(0) \rangle \rangle = A^2 \langle \langle \hat{d}_2^{\dagger}(t) \hat{d}_2(t)
\hat{d}_2^{\dagger}(0)\hat{d}_2(0) \rangle \rangle \nonumber \\
&= A^2\langle \hat{d}_2^{\dagger}(t) \hat{d}_2(0) \rangle\langle \hat{d}_2(t) \hat{d}_2^{\dagger}(0)
\rangle +A^2\langle \hat{d}_2^{\dagger}(t) \hat{d}_2^{\dagger}(0) \rangle\langle \hat{d}_2(t) \hat{d}_2(0) \rangle
\label{Gn2n2}
\end{align}
We retain only connected terms, as indicated by the double angled brackets, and calculate the correlation function in the weak dispersive coupling regime so that the qubit-cavity coupling can be ignored; the remaining system Hamiltonian in terms of the $\hat{d}_j$ operators - given by the sum $\hat{H}_1^d + \hat{H}_2^d + \hat{H}_{12}^d$ - is quadratic [cf. Eqs.~(\ref{H1d}),~(\ref{H2d}),~(\ref{H12d})]. We therefore use Wick's theorem in going from the first line to the second to simplify the average. Fourier transforming the autocorrelation function $G_{F_{\rm Q}F_{\rm Q}}(t)$ allows computing the noise spectral density of the quadratic backaction, $S_{F_{\rm Q}F_{\rm Q}}[\omega]$, following application of the Wiener-Khinchin theorem~\cite{ClerkRMP},
\begin{align}
S_{F_{\rm Q}F_{\rm Q}}[\omega] = A^2\int_{-\infty}^{\infty}\frac{d\omega'}{2\pi}~\Big\{ &S_{d_2^{\dagger}d_2}[\omega']S_{d_2d_2^{\dagger}}[-\omega'-\omega]  \nonumber
\\
+ &S_{d_2^{\dagger}d_2^{\dagger}}[\omega']S_{d_2d_2}[-\omega'-\omega] \Big\} 
\label{Sn2n2}
\end{align}
where the spectral density of two operators $\hat{X},~\hat{Y}$ is simply:
\begin{equation}
S_{XY}[\omega] = \langle \hat{X}[\omega]\hat{Y}[-\omega] \rangle
\end{equation}
Fluctuations in the linear cavity mode $\hat{d}_2$ arise from three types of noise terms; in addition to noise incident directly on the linear cavity (a contribution independent of $\lambda_{12}$), $\hat{d}_2$ is driven by noise incident on the nonlinear cavity that enters the linear cavity by virtue of the cavity-cavity coupling, and is hence $\propto \lambda_{12}$. Lastly, terms $\propto \lambda_{12}^2$ involve noise incident on the linear cavity passing to the nonlinear cavity, being amplified, and then returning to drive $\hat{d}_2$. 

For some analytic insight into the quadratic backaction, we focus on the zero frequency noise $S_{F_{\rm Q}F_{\rm Q}}[0]$. The integrals over frequency space mean that $S_{F_{\rm Q}F_{\rm Q}}[0]$ sees noise at all frequencies (cf. Eq.~(\ref{Sn2n2})), and as a result we expect large contributions near two frequencies, as indicated in Eq.~(\ref{quadraticBA}). Near $\omega = 0$, the nonlinear cavity response $\mathcal{G}[0]$ is the dominant contribution, which also sets the bandwidth $\Omega[\omega = 0] = \kappa_1/\sqrt{\mathcal{G}}$. The linear cavity response here is set by the detuning $\Delta_2$ since the linear cavity is strongly detuned. The contribution to the quadratic backaction near $\omega = 0$ is then:
\begin{align}
S_{F_{\rm Q}F_{\rm Q} }^{(1)} \simeq A^2 \left( \widetilde{\lambda}_{12}^{(1)} \right)^4 \!\! \cdot \frac{1}{\Delta_2^4}  \cdot \frac{\mathcal{G}^2}{\kappa_1^2} \cdot \frac{\kappa_1}{\sqrt{\mathcal{G}}} = A^2 \left( \widetilde{\lambda}_{12}^{(1)} \right)^4 \!\!\cdot \frac{\mathcal{G}^{3/2}}{\kappa_1 \Delta_2^4} 
\end{align}
On the other hand, near $\omega = \Delta_2$, the resonant linear cavity susceptibility dominates the backaction as $|\chi_2| \simeq 1/\kappa_2$, and the bandwidth is therefore $\Omega[\omega = \Delta_2] = \kappa_2$. The nonlinear cavity response here is suppressed by a factor of the large linear cavity detuning, so that $\mathcal{G}[\omega = \Delta_2] \simeq (\kappa_1/\Delta_2)^2$. The quadratic backaction near $\omega = \Delta_2$ is then
\begin{align}
S_{F_{\rm Q}F_{\rm Q}}^{(2)} \simeq A^2\left( \widetilde{\lambda}_{12}^{(1)}\right)^4 \!\!\cdot \frac{1}{\kappa_2^4} \cdot \frac{\kappa_1^2}{\Delta_2^4} \cdot \kappa_2 = A^2\left( \widetilde{\lambda}_{12}^{(1)}\right)^4 \!\!\cdot \frac{\kappa_1^2}{\kappa_2^3\Delta_2^4}
\end{align}
The above estimates for the main contributions to the quadratic backaction integrals (cf. Eq.~(\ref{Sn2n2})) are found to approximate the full integrals well. In the optimization procedure detailed in Section~(\ref{optimalParameters}), these terms are required to be small relative to the intrinsic nonlinear cavity backaction, leading to the constraints in Eq.~(\ref{SFQFQConstraint}).

\begin{figure}[t]
\includegraphics[scale = 0.61]{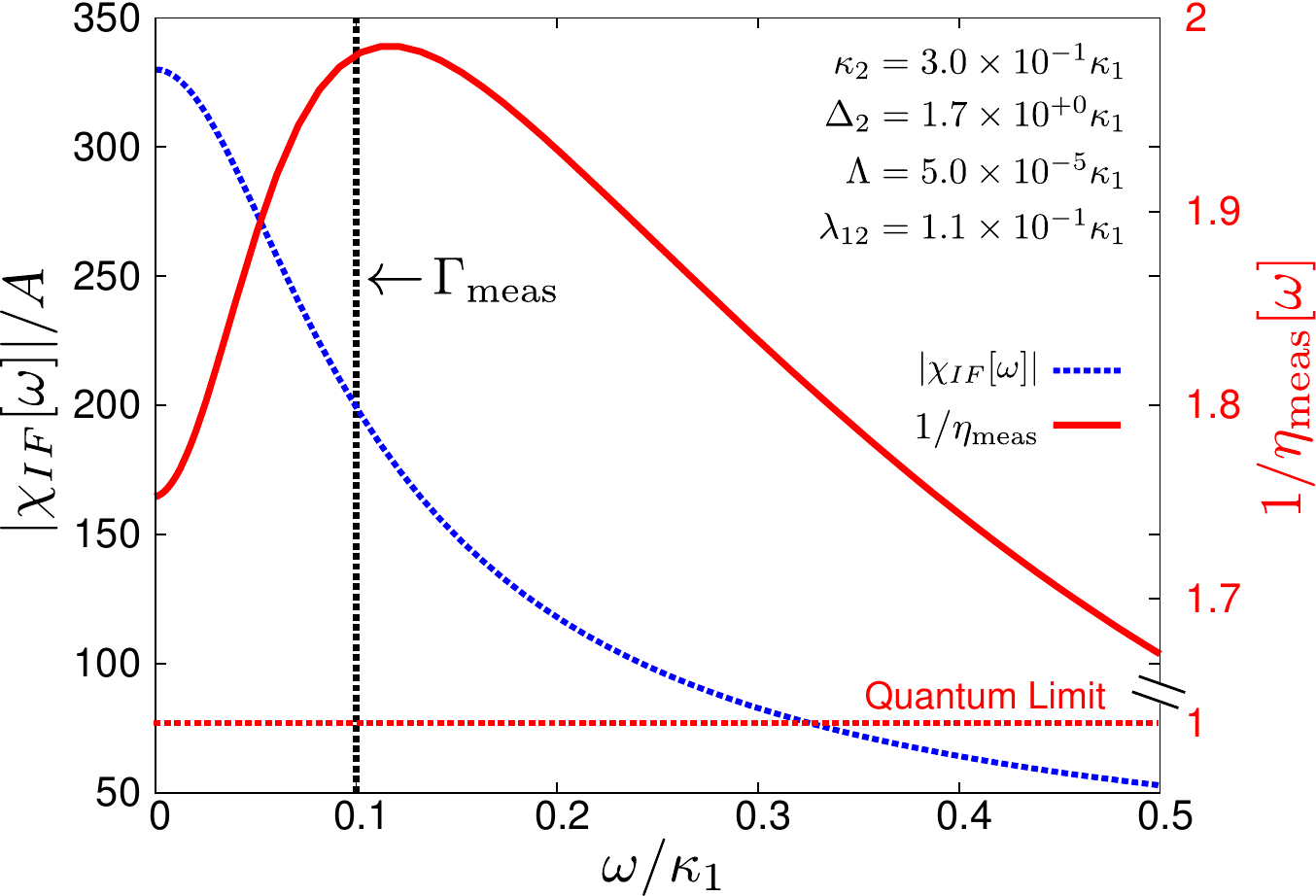}
\caption{Plot of the amplifier forward gain scaled by the dispersive qubit-cavity coupling, $|\chi_{IF}[\omega]|/A$ (dashed blue, left-hand axis) and the inverse efficiency ratio $1/\eta_{\rm meas}$ (solid red, right-hand axis), both as a function of frequency $\omega$, for the same parameter set as Fig.~\ref{GainVsDelta1}. As frequency increases, $1/\eta_{\rm meas}$ first increases slowly as the reduced backaction condition is no longer satisfied. However, the concurrent reduction in gain $\mathcal{G}[\omega]$ leads eventually to a reduced value of $1/\eta_{\rm meas}$, as expected. The vertical dashed line indicates the measurement rate $\sim 0.1~\kappa_1$ for our parameter choices. We see that over this frequency range, $|\chi_{IF}[\omega]|$ decreases somewhat; this is due to $\Gamma_{\rm meas}$ becoming comparable to the amplification bandwidth, a feature that is controlled by the experimental value of $\kappa_1$. Importantly, $1/\eta_{\rm meas}$ increases relative to the zero frequency value by only a small amount, while the forward gain decreases by less than a factor of $2$; our interest in the zero frequency performance of the PARI is therefore justified.}
\label{GainBandwidth}
\end{figure} 


\section{Output homodyne current calculations}
\label{appendixGain}

The full output homodyne current from the nonlinear cavity is the result of interference of the classical homodyne reference beam and the cavity output field; it can be written as~\cite{Gardiner85}:
\begin{equation}
\hat{I}_1[\omega] = \sqrt{\kappa_1}(\cos \phi_h~\hat{x}_1^{\mathrm{out}}[\omega] + \sin \phi_h~\hat{p}_1^{\mathrm{out}}[\omega])
\label{Ihom}
\end{equation}
In standard input-output theory, these cavity output quadratures can be written in terms of the reflected input drives, and the intracavity field leaking out of the cavity~\cite{ClerkRMP}. Using the solutions to Eqs.~(\ref{effSys}) in the rotated basis, the output homodyne current can be written in terms of the amplified and squeezed quadratures. In the particular case of the \emph{average} homodyne current, the vacuum noise drive vanishes, leaving:
\begin{align}
\langle \hat{I}_1[\omega] \rangle = \kappa_1\left[ \cos \phi_h\langle \hat{X}_e[\omega] \rangle +  \sin \phi_h\langle \hat{P}_e[\omega] \rangle \right]
\label{IhomXePe}
\end{align}
where we take $\phi_h \to \phi_h + \theta_e/2$ to absorb angles associated with the transformation.  The cavity drives can be related to the qubit states using the frequency-dependent version of Eq.~(\ref{qubitSub}):
\begin{align}
&\hat{X}_e^{\mathrm{in}}[\omega] \to \hat{X}_e^{\mathrm{in}}[\omega] - \nonumber \\
&\left( \frac{\widetilde{A}~\widetilde{\lambda}_{12}^{(1)}}{\sqrt{2}}\left[i\chi_{2}[\omega]e^{i(\mu_{12} - \nu)} - i\chi_{2}^*[-\omega]e^{i(\mu_{12} - \nu)} \right]\right) \hat{\sigma}_z \nonumber \\
&\hat{P}_e^{\mathrm{in}}[\omega] \to \hat{P}_e^{\mathrm{in}}[\omega] + \nonumber \\
&\left( \frac{\widetilde{A}~\widetilde{\lambda}_{12}^{(1)}}{\sqrt{2}}\left[\chi_{2}[\omega]e^{i(\mu_{12} - \nu)} + \chi_{2}^*[-\omega]e^{i(\mu_{12} - \nu)} \right]\right) \hat{\sigma}_z 
\label{qubitSubFull}
\end{align}
where we suppress the frequency label on $\hat{\sigma}_z$. Following this substitution into Eq.~(\ref{effSys}) and taking the ensemble average, the averaged quadratures $\langle \hat{X}_e[\omega]\rangle,\langle\hat{P}_e[\omega]\rangle$ can be substituted in Eq.~(\ref{IhomXePe}). The amplifier forward gain for arbitrary frequency is given simply by the frequency-dependence version of the linear response relation in Eq.~(\ref{gain}),
\begin{equation}
\chi_{IF}[\omega] =  \frac{\widetilde{A}~\widetilde{\lambda}_{12}^{(1)}}{\sqrt{2} }\kappa_1\sec \theta_e\left( S_1\chi_{e-}[\omega] + S_2\chi_{e+}[\omega] \right)  
\label{gainFull}
\end{equation}
where the coefficients $S_1[\omega], S_2[\omega]$ are given by
\begin{align}
S_1 &= \cos \phi_h \times \nonumber \\
&~~~~-\Big( i\chi_{2}[\omega]e^{i(\mu_{12} - \nu + \theta_e)}  -i\chi_{2}^*[-\omega]e^{-i(\mu_{12} - \nu + \theta_e)} \Big) \nonumber \\
S_2 &= - \sin \phi_h\Big( \chi_{2}[\omega]e^{i(\mu_{12} - \nu)}  + \chi_{2}^*[-\omega]e^{-i(\mu_{12} - \nu)}  \Big)  
\end{align}
Near bifurcation, where the $S_1$ term dominates, the magnitude of the gain reduces to the result in Eq.~(\ref{gain}). Furthermore, in the strongly detuned regime where $\phi_2 \to \pi/2$, and once the reduced backaction condition (cf. Eq.~(\ref{reducedF})) is satisfied, the angular term $\rho_e$ in Eq.~(\ref{gain}) takes the form:
\begin{align}
\rho_e &= \sec \theta_e~\sin \left[ (2M+1)\frac{\pi}{2} + \pi + \theta_e \right]\cos \phi_h \nonumber \\
&\implies \sec \theta_e~\cos \theta_e~\cos \phi_h   = \cos \phi_h
\label{rhoSimple}
\end{align}
Clearly, choosing $\phi_h = 0$ sets $\rho_e \to 1$, allowing us to drop this angular factor. 

The imprecision noise spectral density of the output signal is defined in terms of the full output homodyne current for zero qubit coupling, analogously to $S_{FF}[\omega]$ (cf. Eq.~(\ref{SFFdef})),
\begin{equation}	
	S_{II}[\omega] \equiv \int_{-\infty}^{\infty} dt~e^{i \omega t} \langle \hat{I}(t) \hat{I}(0) \rangle_0.
\label{SIIdef}
\end{equation}
$S_{II}[\omega]$ is straightforward to compute by expressing the full output homodyne current in terms of $\hat{X_e}, \hat{P}_e$, and making use of the usual correlation functions for the input drives [cf. Eqs~(\ref{corrs})]. 

While the complete $\omega$-dependent expressions are unwieldy, we can observe the frequency dependence by computing the inverse efficiency ratio $1/\eta_{\rm meas}$ as a function of frequency for parameter choices indicated in Table~\ref{parChoice}; this is plotted in solid red in Fig.~\ref{GainBandwidth}. We also include the forward gain $|\chi_{IF}|$ (dashed blue). 

\begin{figure}[t]
\includegraphics[scale = 0.7]{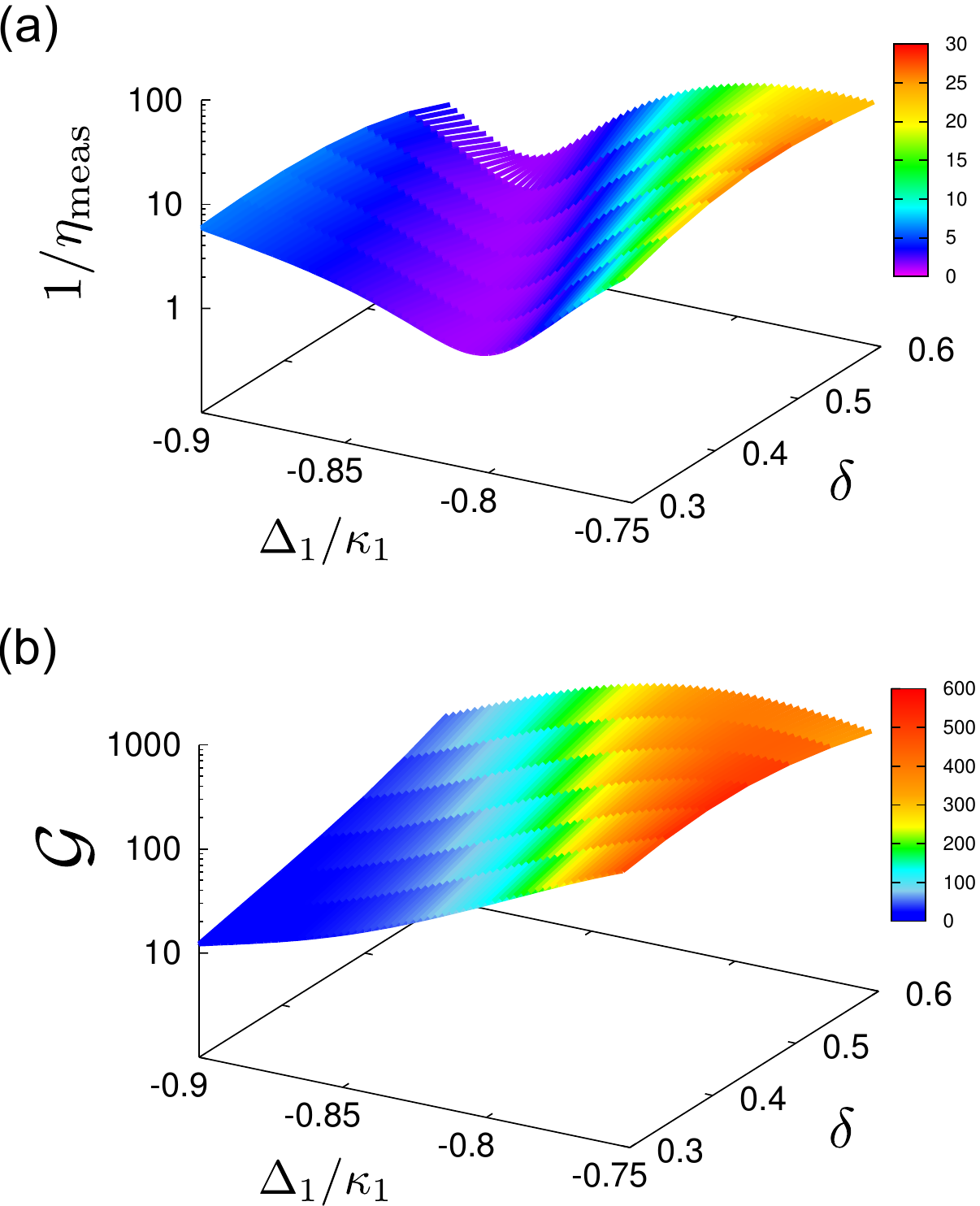}
\caption{3D plot against nonlinear cavity detuning $\Delta_1$ and relative phase difference $\delta$ between cavity drives, of (a) the inverse efficiency ratio $1/\eta_{\rm meas}$. The valley (magenta) indicates a region where $\eta_{\rm meas} \geq 0.5$; (b) the parametric gain $\mathcal{G}$, for the same region as in (a). Clearly, in regions where the PARI performance is nearly quantum-limited [the valley in Fig.~\ref{productSurf}~(a)], $\mathcal{G}$ can be large, of $O(20~{\rm dB})$.}
\label{productSurf}
\end{figure} 


\section{Freedom of parameter choice}
\label{appendixGraphs}

In this section we include supplementary graphs that indicate the freedom of parameter choices afforded by the PARI scheme. \\

Figs.~\ref{productSurf}~(a),~(b) are 3-D plots, against $\Delta_1$ and the cavity drive phase difference $\delta$, of the inverse efficiency ratio $1/\eta_{\rm meas}$ and the parametric gain $\mathcal{G}$ respectively, for fixed nonlinear cavity drive $\bar{a}_1^{\mathrm{in}}$; all other parameters are fixed at the values indicated in Table~\ref{parChoice}. The `valley' (colored magenta) in Fig.~\ref{productSurf}~(a) indicates a region where the deviation from the quantum limit is at most a factor of $2$ ($\eta_{\rm meas} \geq 0.5$); clearly, a nontrivial set of parameters exists where this is true. Fig.~\ref{productSurf}~(b) indicates large gain over parameters corresponding to this valley; therefore, there are multiple parameter sets that are good choices for operation of the PARI detector.  \\

Fig.~\ref{SffProductFinalLambda} is a plot of the gain $\mathcal{G}$ and inverse efficiency ratio $1/\eta_{\rm meas}$ as a function of nonlinear cavity detuning $\Delta_1$ with a Kerr constant $\Lambda =10^{-4}\kappa_1$. This is twice as strong as the nonlinearity used for Figs.~\ref{GainVsDelta1},~\ref{MeasVsDelta1}. We find at the optimal detuning an inverse efficiency of $\simeq 2.01$, with $\mathcal{G}$ still being large, $\mathcal{G} \sim 80$. Strengthening $\Lambda$ reduces the allowed values of $\bar{n}_j$, and hence we pay a cost in terms of the measurement rate; $\Gamma_{\rm meas}$ now is $\sim 1~{\rm MHz}$, a factor of $10$ slower than that in Fig.~\ref{MeasVsDelta1}. Here, the drive parameters are given by: $(\bar{a}_1^{\rm in})^2\!/\kappa_1 = (43.8)^2$, $(\bar{a}_2^{\rm in})^2\!/\kappa_1 = (57.0)^2$, $\bar{n}_1 \sim 3000$, $\bar{n}_2 \sim 700$, and $\delta \simeq 0.06$. Again, an optimization procedure in $\bar{a}_1^{\rm in}$-$\Delta_1$-$\delta$ space has been performed. \\


\begin{figure}[t]
\includegraphics[scale = 0.61]{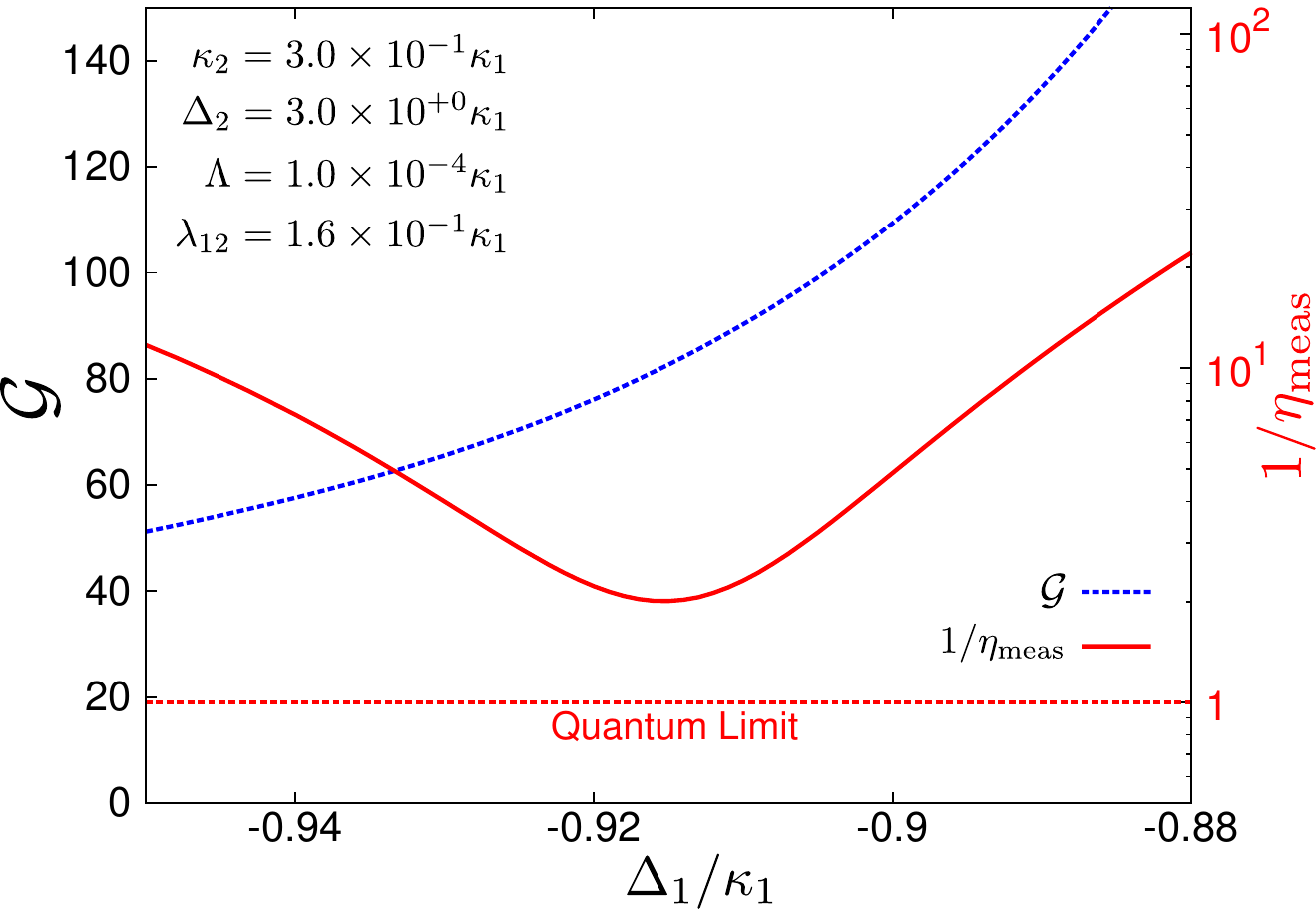}
\caption{Plot of the parametric gain $\mathcal{G}$ (dashed blue, left-hand axis) and the inverse efficiency ratio $1/\eta_{\rm meas}$ (solid red, right-hand log axis) as a function of nonlinear cavity detuning $\Delta_1$, for a stronger $\Lambda$ than in Figs.~\ref{GainVsDelta1},~\ref{MeasVsDelta1}. For optimal detuning, we find $1/\eta_{\rm meas} \simeq 2.01$, with $\mathcal{G} \simeq 80$, and $\Gamma_{\rm meas} \simeq 1~{\rm MHz}$.}
\label{SffProductFinalLambda}
\end{figure}

\section{Minimum useful value of $\lambda_{12}$}
\label{appendixMinL12}

From a description of the output homodyne currents, it is possible to derive the minimum useful value of the intercavity
coupling in the PARI scheme. We consider two types of qubit-cavity detector systems:  the PARI detector, with output homodyne current $\hat{I}_1$, and a linear readout cavity capable of quantum limited qubit state measurement with unit gain, having corresponding output current $\hat{I}_2$. The forms of both currents are as follows:
\begin{align}
\hat{I}_1  &\simeq \widetilde{A}~\hat{\sigma}_z \cdot \sqrt{\kappa_1} \cdot |\chi_2|
\cdot \widetilde{\lambda}_{12}^{(1)}\cdot \chi_{e-}  \nonumber \\
\hat{I}_2 &\simeq \widetilde{A}~\hat{\sigma}_z \cdot \sqrt{\kappa_2} \cdot |\chi_2| 
\end{align}
If $\hat{I}_1$ and $\hat{I}_2$ are equal, then using the nonlinear cavity serves no advantage over the single cavity system;
this condition therefore leads to the minimum useful intercavity coupling strength, given in Eq.~(\ref{l12Min}).



\bibliography{PARI_NJP}

\end{document}